\documentclass[%
 reprint,
 amsmath,amssymb,
 aps,
]{revtex4-2}

\usepackage{arydshln}
\usepackage{environ}
\NewEnviron{ignore}{}{}
\usepackage{multirow}
\usepackage{graphicx}% Include figure files
\usepackage{dcolumn}% Align table columns on decimal point
\usepackage{ulem}
\usepackage{bm}% bold math
\usepackage[colorlinks,linkcolor=blue,citecolor=blue,urlcolor=blue]{hyperref}
\begin{document}

\preprint{}

\title{Magnetostriction, piezomagnetism and domain nucleation in a kagome antiferromagnet}

\author{Qingkai Meng$^{1}$, Jianting Dong$^{1}$, Pan Nie$^{1}$,  Liangcai Xu$^{1}$, Jinhua Wang$^{1}$, Shan Jiang$^{1,2}$, Huakun Zuo$^{1}$, Jia Zhang$^{1}$, Xiaokang Li$^{1,*}$, Zengwei Zhu$^{1,*}$, Leon Balents$^{3,4}$, and Kamran Behnia$^{2,*}$ }

\affiliation{(1) Wuhan National High Magnetic Field Center and School of Physics, Huazhong University of Science and Technology,  Wuhan  430074, China\\
(2) Laboratoire de Physique et d'Etude de Mat\'{e}riaux (CNRS),\\ ESPCI Paris, PSL Research University, 75005 Paris, France\\
(3)Kavli Institute for Theoretical Physics, University of California\\  Santa Barbara, California 93106-4030, USA \\ 
(4) Canadian Institute for Advanced Research, Toronto, Ontario, Canada 
}
\date{\today}

\begin{abstract}
Whenever the elastic energy of a solid depends on magnetic field, there is a magnetostrictive response. Field-linear magnetostriction implies piezomagnetism and vice versa. Here, we show that Mn$_3$Sn, a non-collinear antiferromanget with Weyl nodes, hosts a large and almost perfectly linear magnetostriction even at room temperature. The  longitudinal and transverse magnetostriction, with opposite signs and similar amplitude are restricted to the kagome planes and the out-of-plane response is negligibly small. By studying four different samples with different Mn:Sn ratios, we find a clear correlation between the linear magnetostriction, the spontaneous magnetization and the concentration of Sn vacancies. The recently reported piezomagnetic data fits in our picture. We show that linear magnetostriction and piezomagnetism are both driven by the field-induced in-plane twist of spins. A quantitative account of the experimental data  requires the distortion of the spin texture by Sn vacancies. We find that the field-induced domain nucleation within the hysteresis loop corresponds to a phase transition. Within the hysteresis loop, a concomitant mesoscopic modulation of local strain and spin twist angles, leading to twisto-magnetic stripes, arises as a result of the competition between elastic and magnetic energies. 

\end{abstract}
\maketitle

\begin{figure*}
\includegraphics[width=17cm]{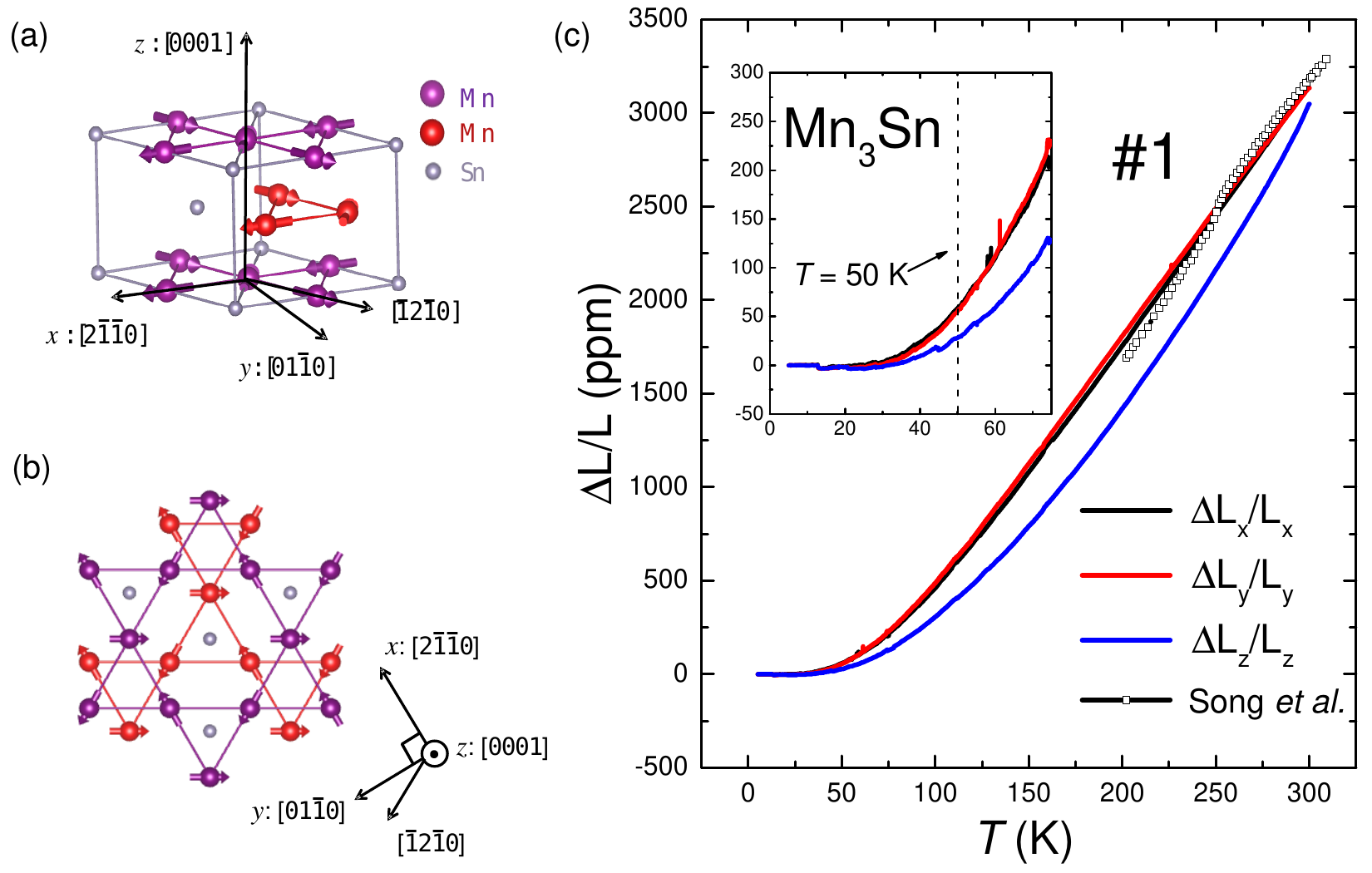}
\caption{\textbf{The magnetic structure and thermal expansion in Mn$_3$Sn}. (a-b). The spin-structure of Mn$_3$Sn, two layers of kagome structure of manganese (a), the Mn atoms in a triangular spin texture rotate 120 degrees (b). The axes called  $x$, $y$ and $z$ axes refer to the [2$\bar{1}\bar{1}$0], [01$\bar{1}$0] and [0001] crystalline orientations. (c) Thermal expansion along three orientations for Mn$_3$Sn comparing to the result of Song \textit{et al.}\cite{Song2018}. The inset shows the thermal expansion responses at low temperatures. The destruction of the inverse triangular spin state indexed by a dashed line, does not cause a detectable anomaly. }
\label{basicMS}
\end{figure*}

\section{Introduction}
Following the discovery of a large room-temperature anomalous Hall effect (AHE) in Mn$_3$Sn \cite{Nakatsuji2015} and in Mn$_3$Ge \cite{Nayak2016}, this family of noncolinear antiferromagnets \cite{Tomiyoshi1982} became a subject of experimental \cite{Li2017,Ikhlas2017, Xu2020,Li2021} and theoretical \cite{Yang2017,Kuroda2017,Liu2017,Miwa2021,Suzuki2017,Zelenskiy2021,Park2018} attention. The AHE signal is present at room temperature and can be inverted with a low magnetic field. Therefore, the discovery has a technological potential in the field of antiferromagnetic spintronics \cite{Tsai2020,Baltz2018,Smejkal2018,Kimata2019,Higo2022nature,Kimata2019}. Very recently, Ikhlas \textit{et al.} \cite{Ikhlas2022Piezomagnetic} reported that Mn$_3$Sn displays a large piezomagnetic effect at room temperature and the sign of the AHE can be modified by a sufficiently large uniaxial stress. By contrasting the response of magnetization and AHE, they demonstrated that the ultimate origin of the AHE is the residual Berry curvature. This confirmed a theoretical prediction \cite{Chen2014}, which argued that AHE in a non-collinear antiferromagnet does not require spontaneous magnetization.

Piezomagnetism, the generation of magnetic moment upon the application of strain, $\sigma$, is intimately linked to magnetostriction, $\epsilon$, the field-induced contraction or elongation of the lattice \cite{Borovik-romanov1994}. Thermodynamics imposes a strict equivalence between the piezomagnetic response ($\frac{\partial M}{\partial \sigma}$) and the field slope of magnetostriction ($\frac{\partial\epsilon}{\partial B}$). Both quantities represent the second derivative of the free energy with respect to magnetic field and stress ($\partial^2F/\partial \sigma\partial B)$ and their unavoidable equality is a specific case of Maxwell relations \cite{pippard1964b}.

Magnetostriction,  first discovered in ferromagnetic iron by Joule \cite{Joule1842}, is technologically attractive \cite{Liu2012MSreview}, because it can be used to conceive devices converting magnetic to mechanical energy and vice versa. Fundamentally, it arises when the magnetic field plays a role in setting the magnitude of the elastic energy. This can happen in metals \cite{Chandrasekhar1971} where it is delectably large (that is $> 10^{-6}$ T$^{-1}$)whenever the electronic density of states depends on magnetic field. Known cases include heavy-fermion systems  near field-induced instabilities \cite{Puech1988} and dilute metals subject to quantizing magnetic fields \cite{Kuchler2014}. In ferromagnets, the dominant magnetostrictive response is due to field-induced displacement and rotation of magnetic domains and the boundary between them. When all domains align and magnetization saturates, magnetostriction collapses \cite{Lee1955,Chopra2015} (See also our data on cobalt below). In antiferromagnets, magnetostriction, drawing significant interest recently\cite{Ma2021,Ikhlas2022Piezomagnetic,Jaime2017}, is usually small and quadratic in magnetic field \cite{Birss1963,Alberts1961,Lines1979}. Save for a limited group of 35 antiferromagnetic classes (out of a total of 122 magnetic point groups), symmetry considerations forbid linear magnetostriction \cite{Birss1963,Tavger1958,Moral1974}.

Here, we report on a detailed study of magnetostriction in Mn$_3$Sn, which belongs with a $mm'm'$ magnetic point group \cite{Suzuki2017}, not of the 35 antiferromagnetic classes allowing linear magnetostriction \cite{Birss1963,Tavger1958,Moral1974}). The linear magnetostriction is allowed by symmetry, because of the residual ferromagnetism. 

By studying four different samples, we find that the slope of linear magnetostriction, $\Lambda$, and the spontaneous magnetization, $M_0$ both depend on the concentration of Sn vacancies. The dependence of $\Lambda$ on Sn concentration  allows us to reconcile the amplitude of the piezomagnetic coefficient resported by Ikhlas \textit{et al.} \cite{Ikhlas2022Piezomagnetic}, with our data. 

We argue that the field-induced in-plane distortion of the spin texture and the competition between magnetic and elastic \cite{Theuss2022} energies leads to the emergence of linear magnetostriction. In this anti-chiral \cite{Balents2022} spin texture, a magnetic field oriented along the kagome planes steadily twists the spins, as seen by torque magnetometry \cite{Li2022torque}. The field-induced rotation of spins generates an almost perfectly linear magnetostriction from 0.02 T to 9 T, which is to be contrasted with the magnetostriction caused by re-orientation of spins close to a ferromagnetic-antiferromagnetic phase boundary \cite{Song2021}. 

We also find that magnetostriction displays a discontinuous jump at the threshold magnetic field for domain nucleation. This is a signature of a second-order phase transition. The amplitude of the jump implies that, during the passage between single-domain regimes of opposite polarities, there are structural stripes  with a mesoscopic width. We dub these stripes twisto-magnetic, since they refer to concomitant modulations of local strain and local spin orientation. Pinning down their structural details emerges as a subject for future experimental and theoretical studies.

\begin{figure*}
\includegraphics[width=17cm]{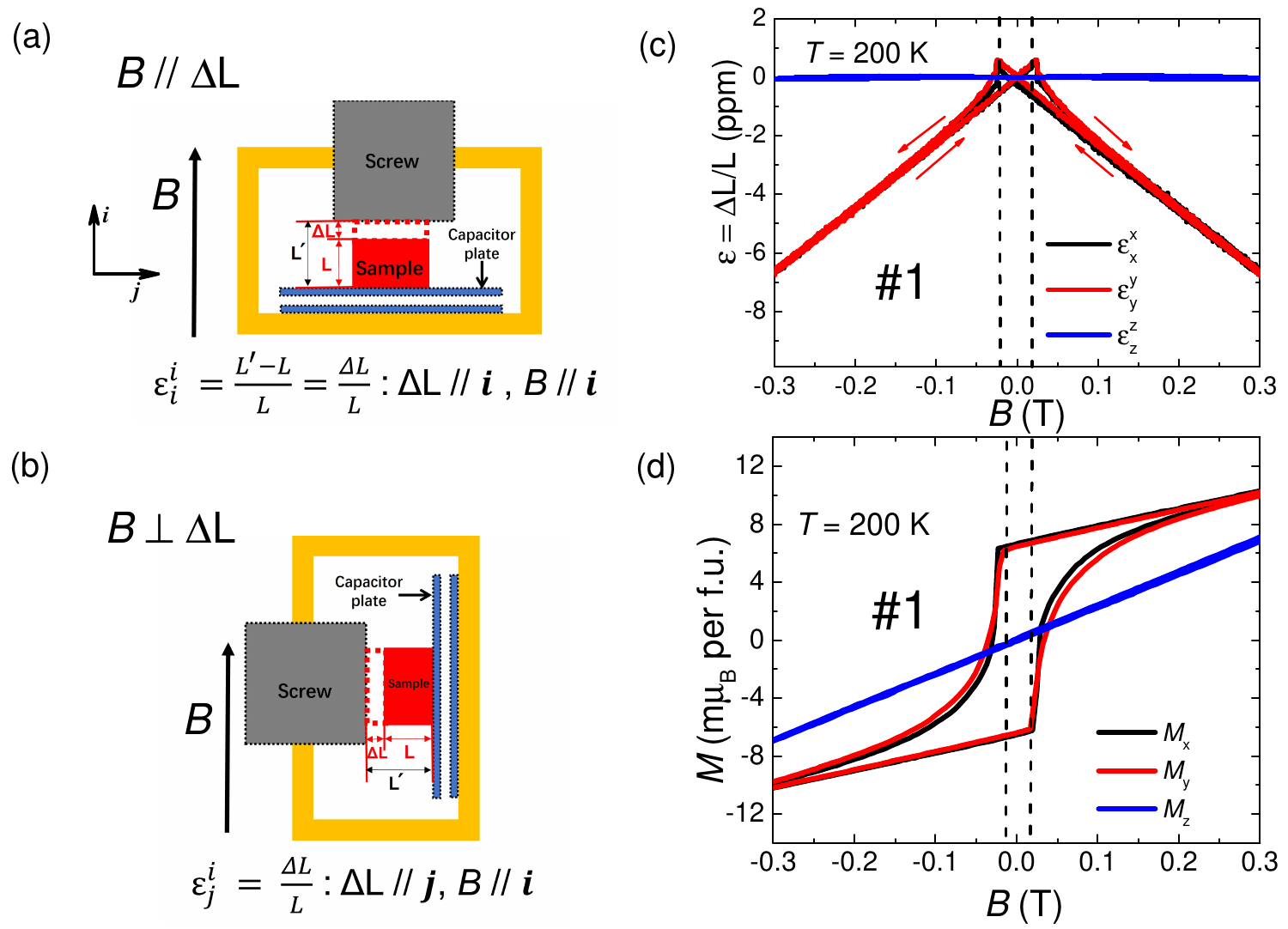}
\caption{\textbf{The magnetostriction in Mn$_3$Sn. } (a-b) The setup to measure longitudinal and transverse magnetostriction. The magnetostriction was measured by the capacitance between two metallic plates. The length change of the clamped sample leads to the change of the distance between the plates, and therefore the capacitance. (c-d) Longitudinal magnetostriction (top) and magnetization (bottom) as a function of the magnetic field for three crystalline orientations. The in-plane magnetostriction ($B$ // $x$, $y$) is large, but for out of plane orientation ($B$ // $z$), it is negligible. Magnetization displays the same slope for the three orientations. Dashed vertical lines indicate the threshold fields for domain inversion }
\label{Setup}
\end{figure*}

\begin{figure*}
\includegraphics[width=17cm]{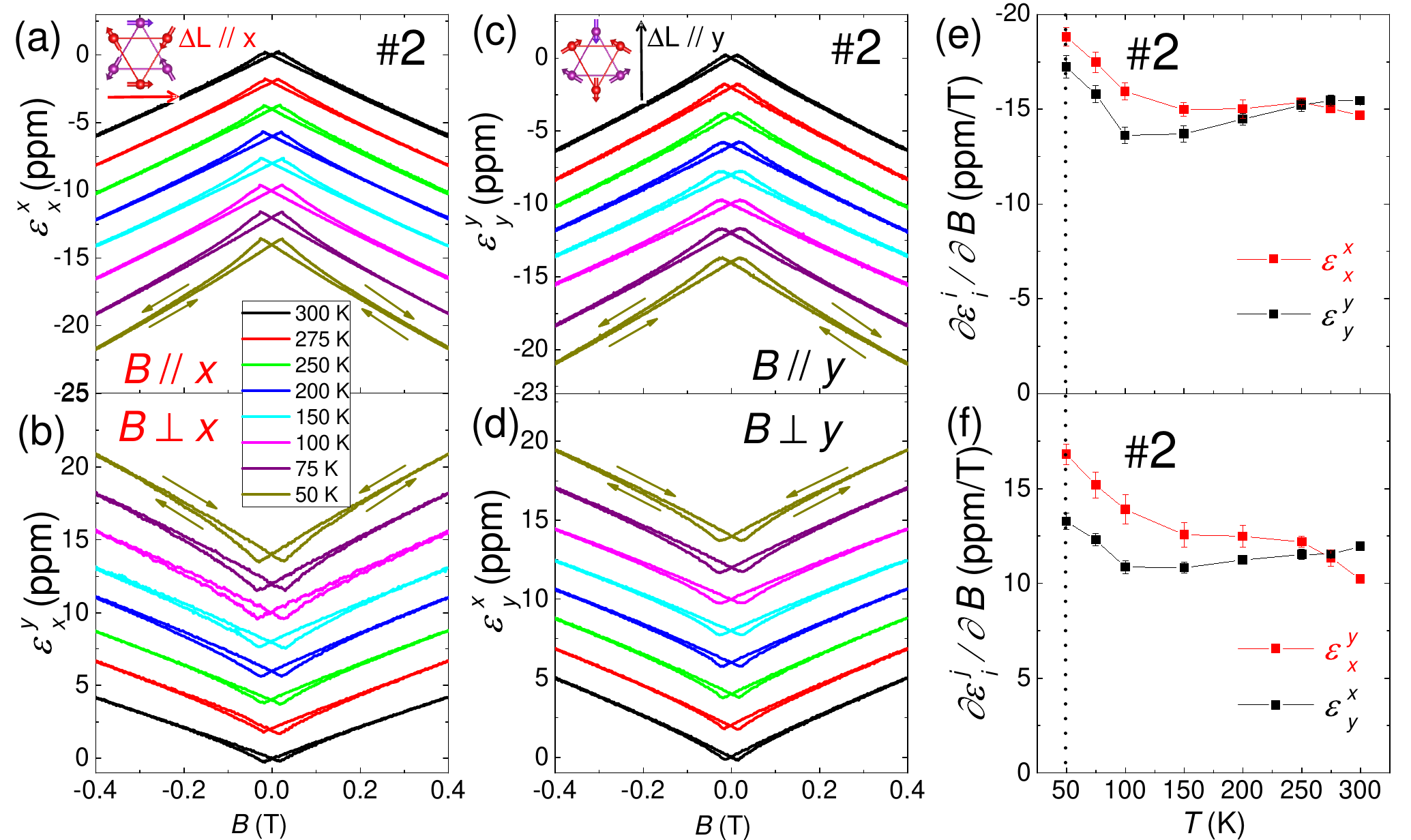}
\caption{\textbf{Longitudinal and transverse magnetostriction for two orientations. } (a-d) The longitudinal and transverse magnetostriction for $\Delta L$//$x$ (a-b) and $\Delta L$//$y$ (c-d) at various temperatures. (e-f) The temperature dependence of the  slope of the magnetostriction: $(\frac{\partial \epsilon}{\partial B})$ for the four configurations. The vertical line marks  the destruction of the anti-chiral magnetic order at 50 K. Note the opposite signs of the longitudinal and transverse responses. }
\label{BothMS}
\end{figure*}

\section{Experimental}
The single crystals  used in this study were grown by the vertical Bridgman method, as detailed in the ref \cite{Li2018}. Mn$_3$Sn  crystallizes into hexagonal DO$_{19}$ structure. It is known that the single crystals are not stoichiometric and the Mn:Sn ratio is larger than 3
 \cite{Ikhlas2020sampleb}. As we will argue below, this is because a small fraction of Sn sites remain unoccupied and therefore the chemical formula becomes Mn$_{3}$Sn$_{1-\delta}$ \cite{KREN1975226}, with a number of physical properties depending on $\delta$. The four samples, which were extensively studied this had different concentrations of Sn vacancies: Mn$_{3}$Sn$_{0.871}$ (\#1), Mn$_{3}$Sn$_{0.891}$ (\#2), Mn$_{3}$Sn$_{0.827}$ (C1) and Mn$_{3}$Sn$_{0.875}$ (C2). The samples dimensions were 0.82$\times$1.02$\times$1.42 mm$^3$ (\#1), 1.01$\times$1.70$\times$2.05 mm$^3$ (\#2), 0.93$\times$1.80$\times$4 mm$^3$ (C1) and 4$\times$0.5$\times$1.8 mm$^3$ (C2). Samples \#1 and \#2 were from one batch, while samples C1 and C2 were from another batch.

The dilatometer used in this study is sketched in Fig. \ref{Setup}(a),(b). It consists of two  metallic plates forming a capacitor. One is kept still, while the other, to which the sample is clamped can move when the length of the sample changes. As the field is swept, this change gives rise to the variation of the capacitance between the two plates \cite{Kuchler2012,Kubler2014}, which can be measured by a capacitance bridge. Please refer to the Methods section for more measurement details.

%A capacitive dilatometer with a resolution of $\Delta L=0.02$ {\AA} made by Innovative Measurement Technology was used to measure magnetostriction in an Oxford Instruments Teslatron PT. We used an Andeen-Hagerling 2550A capacitance bridge for our measurements. By installing the dilatometer perpendicular to the magnetic field, we also measured magnetostriction in the transverse configuration.  The reliability of our set-up was checked by measuring magnetostriction in a cobalt single crystal (see the supplementary materials \cite{SM}).

\section{Results}
 Fig. \ref{basicMS}(a)(b) shows the crystal and magnetic structure of Mn$_3$Sn. Each spin located on a Mn site is oriented 120 degrees off its adjacent spin on the same triangle in a kagome layer. This is an antichiral structure where the rigid clockwise rotation of the triangle with three spins on its vertices would lead to an anti-clockwise rotation of spins on each site \cite{Balents2022}. The magnetic moment on each Mn site is $\sim$ 3 Bohr magnetons, $\mu_{\rm{B}}$. However, the net magnetic moment is only $\sim$ 0.002-0.003 $\mu_{\rm{B}}$ per Mn atom, indicating almost total compensation between moments oriented 120-degree off each other. This inverse triangular spin structure emerges below $T_N=420$ K \cite{Nakatsuji2015,Tomiyoshi1982} and is destroyed below a temperature, which can be tuned by changing the Mn:Sn ratio.
 \cite{Tomiyoshi1982,Nakatsuji2015,Li2017}.  
 
 \textbf{Zero-field thermal expansion-} Fig. \ref{basicMS}(c) shows the thermal expansion data along the three crystal-axis orientations in sample \#1. The thermal expansion coefficient is positive along the three orientations. It is larger in the kagome planes and there is a modest anisotropy which decreases with warming.  Our thermal expansion data is  agreement with the results of a previous study restricted to temperatures above  200 K \cite{Song2018}. In  Fig. \ref{basicMS}(c), for each orientation, the length at 2 K is taken as the reference. The inset of the Fig. \ref{basicMS}(c) shows the thermal expansion coefficients at low temperatures. The destruction of the inverse triangular spin state around 50 K, indexed by a dashed line, does not cause any detectable anomaly.

 \textbf{Longitudinal and transverse magnetostriction-} Fig. \ref{Setup}(c) compares the longitudinal magnetostriction along the three crystalline orientations in low fields up to 0.3 T at 200 K. In our convention, $\epsilon^i_j$ refers to the magnetostriction measured along $j$ when the magnetic field is along $i$. Longitudinal magnetostriction is contractile, and  almost identical along the $x$ and $y$ axes. On the other hand, when the field is along the $z$-axis, there is no detectable signal. Fig. \ref{Setup}(d) shows the magnetization in the same sample. There is a finite spontaneous magnetization for the two in-plane configurations, but not for the out-of-plane one. Magnetization has a finite  slope along the three orientations, as reported previously \cite{Nakatsuji2015}. The absence of out-of-plane longitudinal magnetostriction implies that uni-axial stress along $z$-axis does not affect out-of-plane magnetization. In contrast, the in-plane longitudinal magnetostriction is finite, implying that in-plane stress would affect the magnetization. Moreover, since the magnetostriction is linear in magnetic field,  stress should shift magnetization without changing its slope as a function of magnetic field. This is indeed what was found by the recent study of piezomagnetism \cite{Ikhlas2022Piezomagnetic}.

As seen in Fig. \ref{Setup}(c)(d),  magnetostriction and magnetization behave concomitantly. When the magnetic field exceeds a critical threshold ($B_0 = 0.02$ T marked by two dashed vertical lines  for the two sweeping orientations), magnetostriction peaks and magnetization starts to increase.  $B_0$ is the field at which domains (with a polarity set by magnetic field and inverse to the prevailing one) nucleate. When the magnetic field exceeds a much larger amplitude  ($\approx 0.2$ T), the system becomes single domain again \cite{Li2018}. We will discuss what happens when $B = B_0$, in more detail later below.

\begin{table*}
\centering
\begin{tabular}{|c|c|c|c|c|c|c|c|c|}
\cline{1-9}
\multirow{2}{*}{Sample}& \multicolumn{2}{c|}{$\epsilon_x^x$}& \multicolumn{2}{c|} {$\epsilon_x^y$} &\multicolumn{2}{c|} {$\epsilon_y^y$ } & \multicolumn{2}{c|} {$\epsilon_y^x$}\\
\cline{2-9}
 & $a_l$&$a_q$ & $a_l$&$a_q$ & $a_l$&$a_q$ & $a_l$&$a_q$ \\
\cline{1-9}
\#1& -1.949& -0.0015 & 0.612& 0.0004&-2.244& 0.0156& 1.091& 0.0012 \\
\cline{1-9}
\#2& -1.297& 0.007 &1.105& -0.0095& -1.174& -0.0092& 1.021& -0.0252 \\
\cline{1-9}
\end{tabular}
\caption{\textbf{Parameters extracted by fitting  the field dependence of the magnetostriction up to 9 T at 200 K for two samples.} The fitting parameters $a_l$ and $a_q$ were obtained by fitting the data to $\epsilon=a_lB+a_qB^2$. The units of  $a_l$ and $a_q$ are $\times$10$^{-5}$ T$^{-1}$ and $\times$10$^{-5}$ T$^{-2}$, respectively. The linear magnetostriction, $\Lambda$ corresponds to $a_l$. One can see it totally dominates the quadratic term, even at the order of 10 T, $a_l \gg a_q\times B$.  }
\label{table-fitting}
\end{table*}

Fig. \ref{BothMS} (a)-(d) show the longitudinal and the transverse  magnetostriction for both $x$ and $y$ orientations at various temperatures. For both orientations, longitudinal magnetostriction is contractile and transverse magnetostriction is expansive. As seen from the figures, it is linear in all four configurations. The field slope of  $\frac{\partial \epsilon}{\partial B}$ is extracted from this data and is plotted as a function of temperature in Fig. \ref{BothMS} (e)-(f). For all four configurations, $\frac{\partial \epsilon}{\partial B}$  are flat above 150 K and begin to increase with decreasing temperature with the approach of the 50 K phase transition (for a brief glimpse to the magnetostriction below 50K, see supplementary Figure 11 in the supplementary materials\cite{SM}). 

We conclude that even at room temperature, for both $x$ and $y$ directions, there is a large longitudinal $\frac{\partial \epsilon}{\partial B}$  of comparable magnitude ($\approx -1.5 \times 10^{-5}$ T$^{-1}$) slightly larger than a transverse signal of opposite sign ($\approx +1.1 \times 10^{-5}$ T$^{-1}$). We also measured transverse magnetostriction for other configurations and found a tiny response within the margin of our experimental resolution (see supplementary Figure 3 in the supplementary materials \cite{SM}). 

\begin{figure}[ht]
\includegraphics[width=8.5cm]{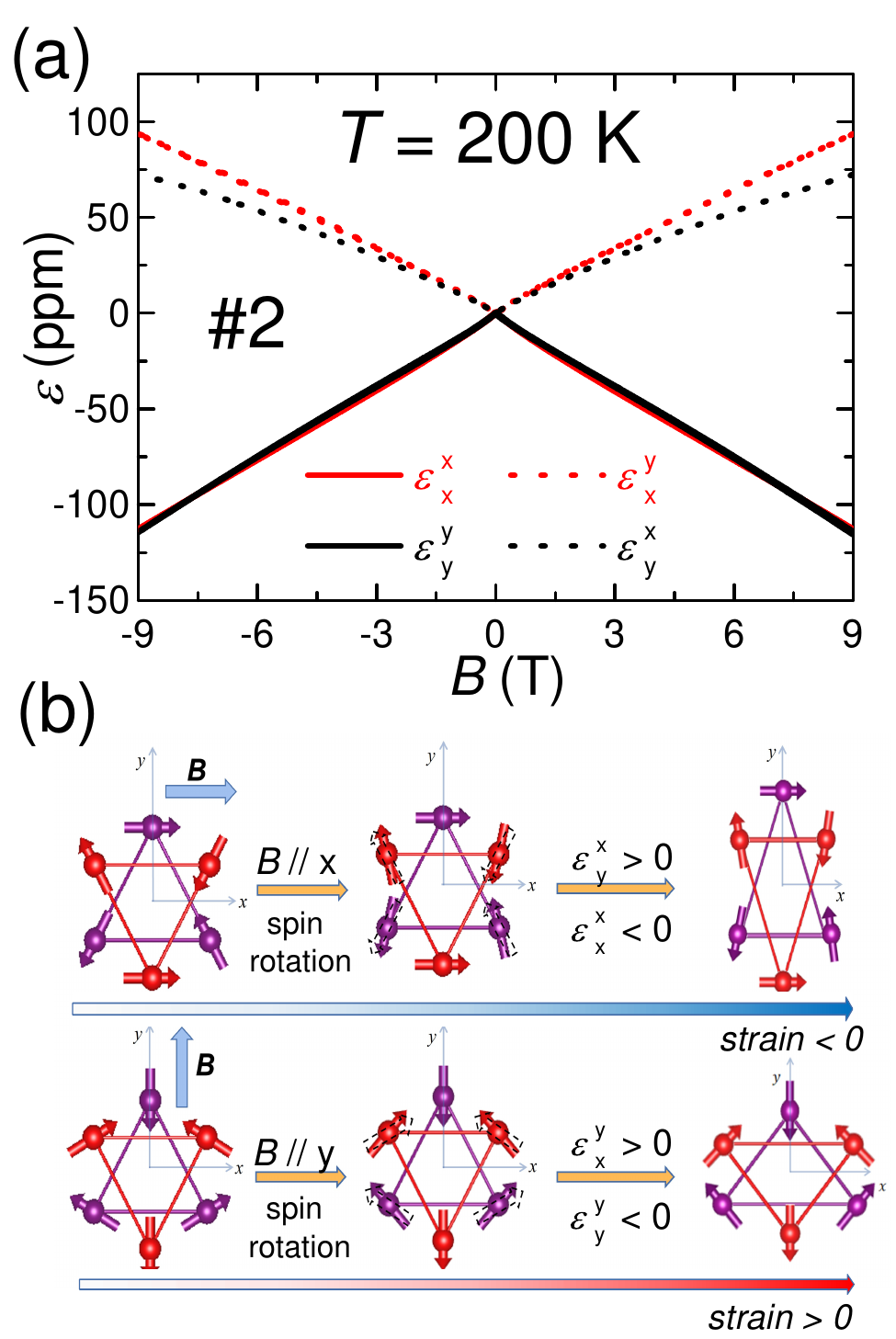}
\caption{\textbf{High-field magnetostriction and the mechanism.} (a) Magnetostriction up to 9 T for  four configurations. The nearly perfect linearity persists as the field is increased. (b) A sketch of the mechanism. When the field is applied along the $x-$ or $y-$ axes, the  spin texture is distorted. Because of the spin-lattice coupling, the distortion of the spin angles leads to  (a much milder) distortion of atomic bonds. }
\label{highB}
\end{figure}

\begin{figure*}[ht]
\includegraphics[width=17cm]{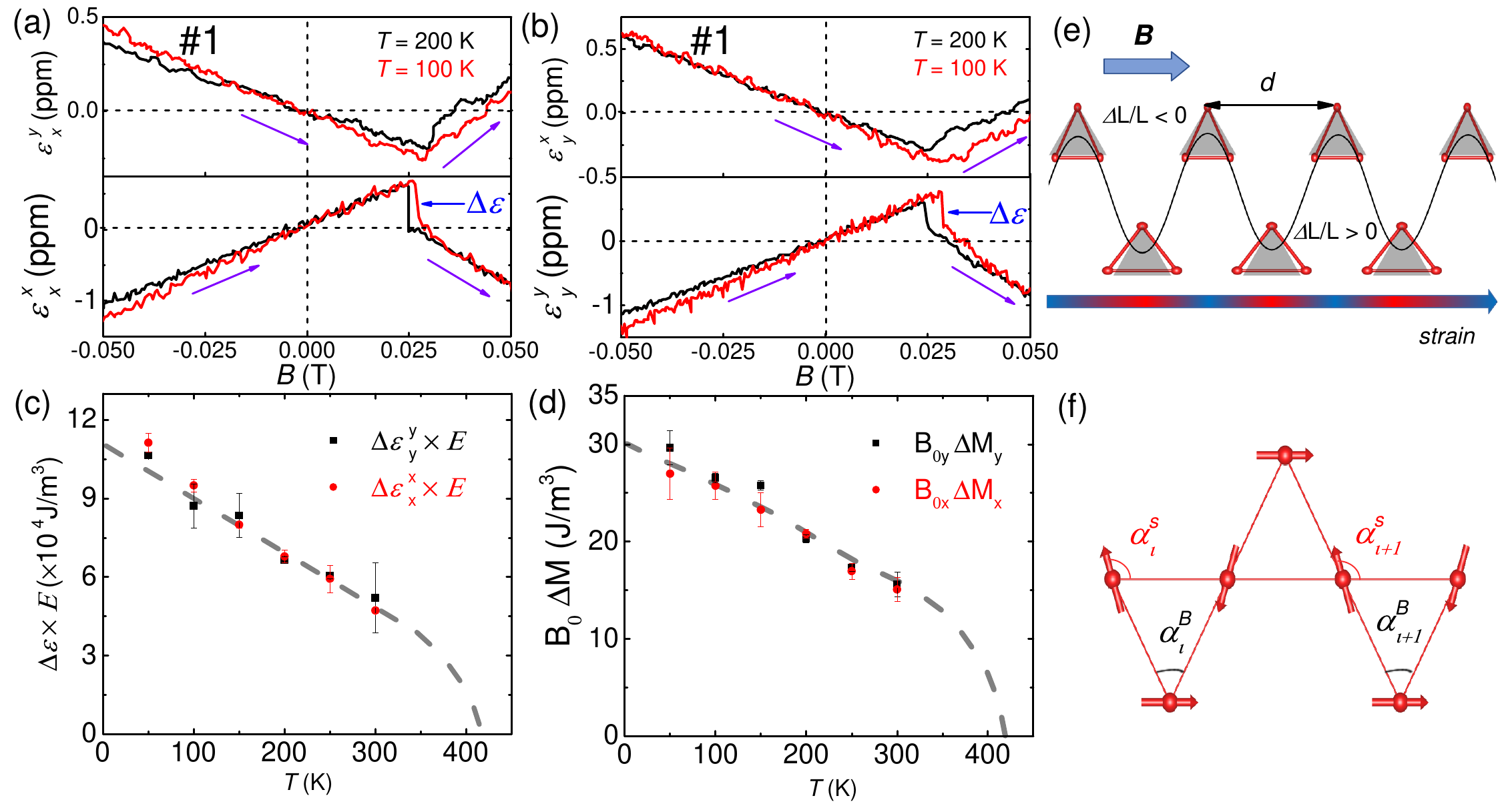}
\caption{\textbf{Discontinuity in a second-order derivative of the free energy-} (a) Transverse and longitudinal magnetostriction for $\Delta L$ // $x$. (b) Transverse and longitudinal magnetostriction for $\Delta L$ // $y$. In both cases, the longitudinal magnetostriction shows a discontinuity at $B_0$. The two vertical dashed lines show the amplitude of magnetostriction because of the domain reversal. (c) The  amplitude of this jump multiplied by the Young Modulus ($E$ = 123 GPa) as a function of temperature. This quantifies the elastic energy associated with the jump. (d) The product of $B_0$ and magnetization as a function of temperature. This quantifies the energy cost of domain nucleation. The two energies differ in amplitude by more than three order of magnitudes. But, both tend to vanish at $T_N$ = 420 K. (e) The sketch of modulation in space at $B_0$: Since  $\epsilon$($B = B_0$) $= \epsilon$($B$ = 0), the overall magnetostriction should be zero with peaks of positive  magnetostriction and valleys of negative magnetostriction, separated by a distance, $d$, much longer than the lattice parameter. (f) The sketch of two neighboring atomic triangles. If the difference in spins' angles ($\alpha_l^s$) becomes much larger than the difference in the atomic bond's angle ($\alpha_l^B$), then the gradient of magnetization is large enough to compensate the mechanical force due to  a gradient in elastic energy.}
\label{lowB}
\end{figure*}
\begin{figure*}[ht]
\includegraphics[width=17cm]{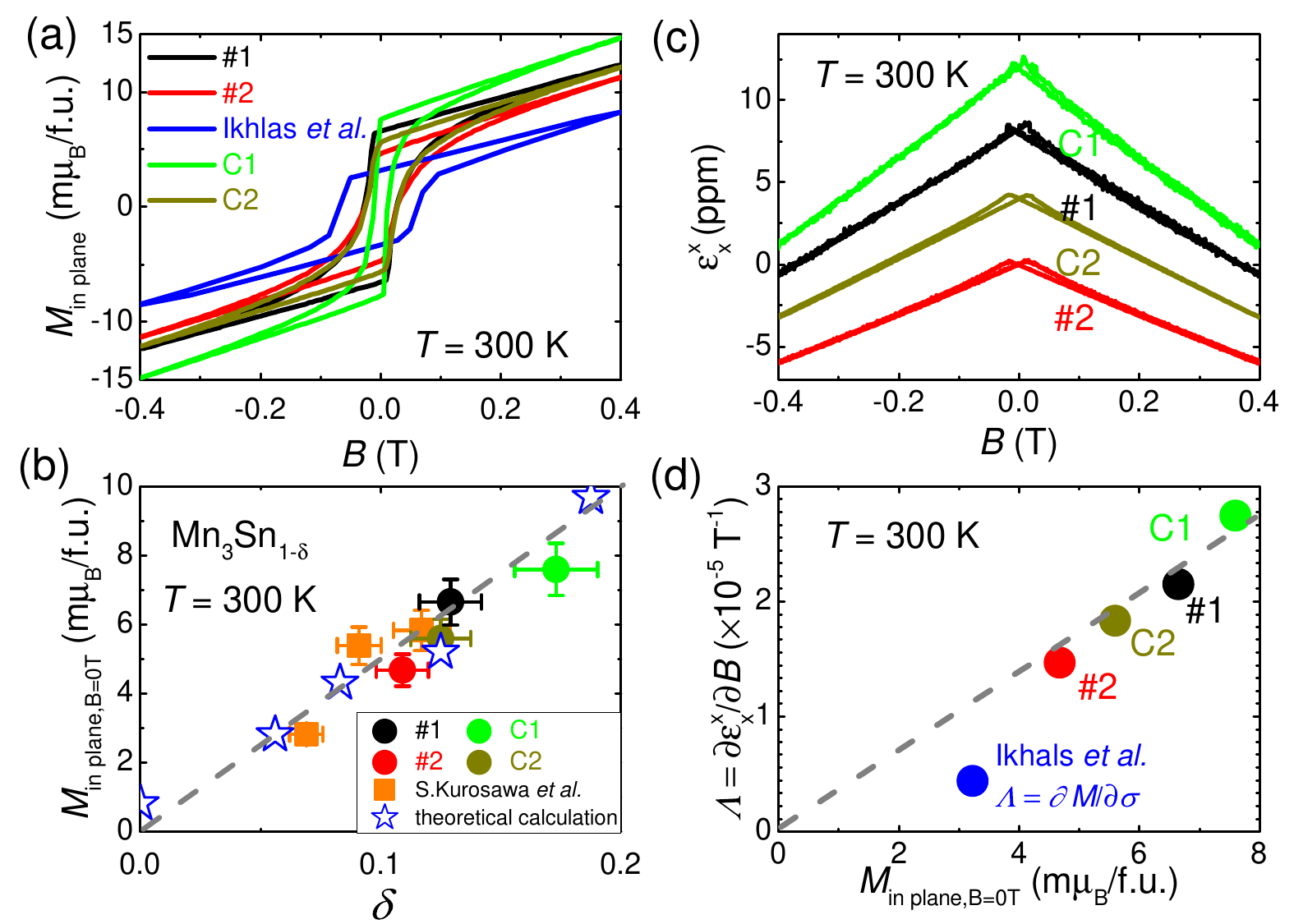}
\caption{\textbf{Amplitude of magnetostriction and magnetization and its correlation with the concentration of Sn vacancies.} (a) Magnetization of different samples at 300 K. (b) The amplitude of the spontaneous magnetization as a function of the concentration of Sn vacancies in our samples and those reported in ref. (c) Magnetostriction of our samples at 300 K.  (d) $\Lambda$ as a function of spontaneous magnetization in our samples and the one reported by Ikhlas \textit{et al.}\cite{Ikhlas2022Piezomagnetic}  at 300 K.  }

 \label{different samples}
 \end{figure*}

\textbf{Linear magnetostriction up to 9 T-} Fig. \ref{highB}(a) illustrates the evolution of magnetostriction up to 9 T. One can express the magnetostriction as the sum of a linear and a quadratic term : $\epsilon=a_lB+ a_qB^2$. Table \ref{table-fitting} lists the best fits to these coefficients for two different samples. 

The combination of a dominant linear ($a_l$) and  a sub-dominant quadratic ($a_q$) term up to $\approx $ 9 T is reminiscent of the case of UO$_2$ in its magnetically ordered state \cite{Jaime2017}. Interestingly, the amplitude  of the linear magnetostriction is roughly similar in the two cases ($\approx 10^{-5}$ T$^{-1}$). However, here it is a room-temperature phenomenon, while in UO$_2$ it emerges below a N\'eel temperature of 30 K\cite{Jaime2017}. No saturation is visible in our data up to 9 T in contrast with what is observed in ferromagnets (For an example, see the case of Co in the supplementary materials \cite{SM}). Fig. \ref{highB}(b) shows a sketch of mechanism. We will discuss the microscopic origin of this robust field-linear magnetostriction in the next section.

\textbf{Discontinuity at the threshold of domain nucleation-} Fig. \ref{lowB}(a),(b), is a zoom on the low-field behavior near $B_0$. One can see a discontinuity in longitudinal magnetostriction at $B_0\approx 0.02$ T. This is the field at which magnetic domains nucleate. When the magnetic field is swept down to zero, the magnetostriction linearly decreases to a null value. It changes sign with the inversion of the magnetic field. But the magnetization and the anomalous Hall effect do not change sign when the magnetic field inverts. They continue to have a sign opposite to the magnetic field. At $B_0$, both the magnetization and the AHE begin to evolve along the orientation imposed by the magnetic field \cite{Li2018}. At this field, longitudinal magnetostriction shows a discontinuous jump (Fig. \ref{lowB}(a),(b)), in contrast with the weaker anomaly seen in transverse magnetostriction. The observed jump represents a discontinuity in the second derivative of the free energy and implies a second-order phase transition. 

The magnetostriction hysteresis loop in Mn$_3$Sn differs qualitatively from other cases of linear magnetostriction such as MnF$_2$ or UO$_2$, where the response is set by domain rotation (see supplementary Figure 6 in the supplementary materials \cite{SM}). We will come back to the implications of this observation and also the Fig. \ref{lowB}(e),(f) in the next section. 

Multiplying the amplitude of the jump in magnetostriction by the Young Modulus ($E$ = 123 GPa \cite{Theuss2022}) allows to quantify the change in elastic energy,  shown in the Fig.\ref{lowB}(c). Similarly, we can identify  the energy cost of domain nucleation with the product of $B_0$ and magnetization, shown in the Fig.\ref{lowB}(d). The two energies differ by more than three orders of magnitudes and both tend to vanish at $T_N$ = 420 K as expected.

\textbf{Sample dependence-}  Having found that the amplitude of linear magnetostriction differs from one sample to the other, we undertook an extensive examination of magnetization and magnetostriction of additional samples. Fig.\ref{different samples}(a) portrays the magnetization of several samples at 300K. In addition to four different samples used in our study, the figure includes the sample studied by Ikhlas \textit{et al.} in their piezomagnetic study \cite{Ikhlas2022Piezomagnetic}.  One can see that the amplitude of the jump in magnetization, $M_0$ varies from sample to sample. 

Fig.\ref{different samples}(b) plots the amplitude of this spontaneous magnetization as a function of $\delta$, the concentration of Sn vacancies. In addition to our samples, we include three samples studied by Kurosawa \textit{et al.} \cite{kurosawa2022chiral}. There is a clear correlation between the amplitude of $M_0$ and $\delta$. This implies that in-plane ferromagnetism is, at least partially, caused by the presence of  Sn vacancies. 

Fig.\ref{different samples}(c) compares the magnetostriction of four different samples. There is a visible difference in slope of $\epsilon^x_{x}$. In other words, linear magnetostriction has not the same amplitude. Plotting  $\Lambda= \frac{\partial}{\partial B} \epsilon^x_{x}$ as a function of $M_0$ ( Fig.\ref{different samples}(d)) of the four crystals shows a clear quasi-linear  correlation between the two. The figure includes another data point, $\Lambda$ reported by the piezomagnetic study \cite{Ikhlas2022Piezomagnetic}.  Thus, residual ferromagnetism is not only indispensable for the existence of linear magnetostriction, but also sets its amplitude. As seen in supplementary Figure 12\cite{SM}, the presence of Sn vacancies leads to a detectable change in the lattice parameter along the $a$-axis. Along the $c$-axis, the variation is undetectably small. Note that the difference in lattice parameter among the samples is at least an order of magnitude larger than the change induced by a 10 T magnetic field in  each sample. 

\section{Discussion}

\textbf{Origin of the linear magnetostriction-} Both piezomagnetism and magnetostriction arise because any
deviation of the elementary units of Mn$_{3}$Sn from a perfect equilateral geometry  disrupts the cancellation of the net moment of the three spins therein. Microscopically, strain transforms the equilateral triangle to an isosceles one. The cancellation is spoiled because the two most nearby spins have a stronger antiferromagnetic coupling, and hence twist slightly toward a more antiparallel alignment.  Consequently, a distortion of the triangle induces a net moment (piezomagnetism), and a net moment caused by field-induced twisting of the spins\cite{Li2022torque,Balents2022}  induces a distortion (magnetostriction).  

In the ideal system, where all triangles are identical, one can understand this effect very simply from both a microscopic calculation and a symmetry point of view.  In this case, we can consider the spins on a single triangle, and define three linear combinations:

\begin{eqnarray}
  \label{eq:39}
   \label{eq:20}
  \bm{M} & =&  \bm{S}_0+\bm{S}_1+\bm{S}_2, \nonumber \\
  \bm{\Phi} & = & \bm{S}_0 + e^{-2\pi i/3}
              \bm{S}_1+e^{-4\pi i/3}\bm{S}_2,
\end{eqnarray}

Here, $\bm{M}$ is a real vector representing the total spin on a triangle, and $\bm{\Phi}$ is a complex vector which defines the antiferromagnetic order parameter.  For spins of fixed length $S$, we have $|\bm{M}|^2 +2\bm{\Phi}^*\cdot \bm{\Phi} = 9 S^2$.   When the three spins are at 120 degrees to one another, $\bm{M}=0$ and $\bm{\Phi}^*\cdot \bm{\Phi} = \frac92 S^2$.  Note that when all spins rotate by 180 degrees,  $\bm{\Phi}$ changes sign, but keeps its absolute value.

Nearest-neighbor exchange in these variables becomes
\begin{equation}
  \label{eq:40}
  J \left(\bm{S}_0\cdot\bm{S}_1+
    \bm{S}_1\cdot\bm{S}_2+\bm{S}_2\cdot\bm{S}_0 \right)  =
  \frac{J}{3}\left(
    |\bm{M}|^2 - \bm{\Phi}^*\cdot\bm{\Phi}\right).
\end{equation}
One can see readily this favors the antiferromagnetic state.  Now we include the simplest exchange-striction effect, which corresponds to the modification of the exchange $J$ on a bond proportional to the change in its length.  This is modeled by the term
\begin{equation}
  \label{eq:41}
  H_{e-s} = -\sum_n J g \left( \bm{\hat{f}}_n\cdot \bm{\varepsilon} \cdot
    \bm{\hat{f}}_n\right) \bm{S}_n \cdot \bm{S}_{n+1},
\end{equation}
where $\bm{S}_3=\bm{S}_0$, $\bm{\varepsilon}$ is the strain tensor,
and $\bm{\hat{f}}_n$ is the unit vector along the bond connecting spin
$n$ and $n+1$, and $g$ is proportional to $-\frac{r}{J}\frac{dJ}{dr}$.    Using Eq.~\eqref{eq:39}, one obtains, dropping a
trivial volume expansion contribution,
 \begin{equation}
  \label{eq:42}
  H_{e-s}  = -\frac{J g }{3}\textrm{Re}\,\left[\varepsilon_+\left(\bm{M} \cdot \bm{\Phi}^*
       - \bm{\Phi}\cdot \bm{\Phi}\right) \right],
 \end{equation}
 where
 $\varepsilon_+ = \frac12 (\epsilon_{xx} - \epsilon_{yy} + 2 i
 \epsilon_{xy})$.

In a pure spiral state (either chiral or antichiral), the $\bm{\Phi}\cdot \bm{\Phi}=0$ term vanishes.  The  remaining term in Eq.~\eqref{eq:42} describes an
 in-plane anisotropic stress arising in the presence of magnetization
 and antiferromagnetic order.  The stress (and consequent strain) is
 induced {\emph {without spin-orbit coupling}}, because the
 antiferromagnet order itself couples spin and orbital symmetries.
 Specifically, a 120 degree ordered state breaks both spin rotation
 and $C_3$ spatial rotation symmetries, but preserves the combination
 of the two -- in the anti-chiral case, the two rotations are made in
 the opposite sense.  One might think of the antiferromagnet itself as a state in which spin-orbit coupling is generated spontaneously!

To understand how this effect arises without SOC, we need to think about the presence of both magnetization and AF order.  In Mn$_3$Sn, the AF order is long-range and well-established at room temperature.  When we think about any single triangle of spins, the AF order parameter $\bm{\Phi}$ is therefore imposed upon it by other triangles coupled to it, and related to the order far away. Supposing this global AF order is somehow pinned (typically by an applied field), then the spin orientations on the triangle in question are almost fixed.  If a small uniform magnetization is imposed, the spins rotate slightly, and then in turn  leads to a difference of bond energies, which depends on the relative orientation of the AF order and the magnetization.  That relative orientation determines the axes for the anisotropic stress, which
 will induce a strain that seeks to strengthen the exchange on the bond with the most anti-aligned pair of spins.

While this effect does not require SOC, other effects can occur when
 SOC is present.  To allow for all possibilities, we consider the
 general form of linear couplings of an anisotropic in-plane strain
 $\varepsilon_+$ to the magnetic order parameters.  Here we specialize
 to the {\em anti}-chiral state, whose order parameter is the complex
 {\em scalar} $\Phi = \Phi_x - i \Phi_y$.  We assume
 $\Phi_z = \Phi_x + i \Phi_y=0$, so there is pure anti-chiral order.
 We also express the in-plane magnetization as a second complex scalar
 $M=M_x-iM_y$.  Then the most general free energy density linear in
 $\varepsilon_+$ is
 \begin{equation}
   \label{eq:43}
   f_{e-s}^{\rm gen} = - \textrm{Re}\, \left[ \varepsilon_+\left(
       \gamma_1 M \Phi^* + \gamma_2 M^2 + \gamma_3 \left(\Phi^*\right)^2\right)\right],
 \end{equation}
 where $\gamma_{1,2,3}$ are phenomenological coupling constants.  By
 comparing to Eq.~\eqref{eq:42}, we can see that $\gamma_1 = \frac{g J
 }{3v_{\rm u.c.}}$, where $v_{\rm u.c.}$ is the volume per unit cell
 used to convert to an energy density.  The other two couplings
 $\gamma_2,\gamma_3$ are zero in the absence of SOC, but are generally
 present and can be obtained for example from single ion anisotropy (SIA) \cite{Ikhlas2022Piezomagnetic}.  

Using the equations of elasticity, we have $\bm{\varepsilon} = \bm{C}^{-1} \bm{\sigma}$, where $\bm{\sigma} = - \partial f_{\rm
  e-s}/\partial \bm{\varepsilon}$ is the strain induced by the
spin-lattice coupling.  For the case of polar anisotropy
and a three-fold rotation axis along $z$, we find
 \begin{equation}
   \label{eq:44}
   \epsilon_+ = \frac{1}{2 C_{66}} \left(\gamma_1 M^* \Phi + \gamma_2 \left(M^*\right)^2 + \gamma_3 \Phi^2\right),
 \end{equation}
where $C_{66}$ is the shear modulus in Voigt notation.  For fields
which are larger than the hysteresis field, but still small enough
that the magnetization is small, one can approximate $M
\approx M_s + \chi H$, and $\Phi \sim |\Phi| \hat{H}^*$, where $H = H_x -
i H_y$ and $\hat{H} = H/|H|$.  We then have
\begin{eqnarray}
  \label{eq:45}
  \epsilon_+ & \approx & \frac{1}{2 C_{66}} \left(\gamma_1 |\Phi| M_s^*
    \hat{H}^* + \gamma_2 \left(M_s^*\right)^2 + \gamma_3|\Phi|^2
    (\hat{H}^*)^2\right) \nonumber \\
    &+&   \frac{\chi}{2 C_{66}}
  \left(\gamma_1 |\Phi|\hat{H}^*+ 2\gamma_2  M_s^*
      \right)H^* + O(H^2).
\end{eqnarray}
The second line describes linear in field magnetostriction:
\begin{eqnarray}
  \label{eq:46}
 \frac{d\epsilon_+}{dH^*} & \approx &\frac{\chi}{2 C_{66}}
  \left(\gamma_1 |\Phi|\hat{H}^*+ 2\gamma_2  M_s^* \right)
\end{eqnarray}

We see that it arises from two contributions.  The ``intrinsic'' term
obtained above gives $\gamma_1$, which is independent of the
spontaneous magnetization $M_s$.  By contrast, the ``extrinsic'' term
$\gamma_2$ gives a linear contribution proportional to $|M_s|$.

Consistency with the experiment would be achieved if $\gamma_2$ were
the dominant contribution.  This is surprising and in disagreement
with na\"ive expectations, and $\gamma_2 \ll \gamma_1$ for the uniform
system in the weak anisotropy limit.  It seems that non-stoichiometry
has the effect of dramatically enhancing $\gamma_2$.

\textbf{Neumann's principle-} From a symmetry point of view, linear magnetostriction may occur in certain antiferromagnetic point groups\cite{Birss1963}, but occurs generically in ferromagnets.  While Mn$_3$Sn is dominated by antiferromagnetism, it is in the symmetry sense a ferromagnet: it belongs to the ferromagnetic point group  $mm'm'$ (see the supplementary materials \cite{SM}).  Consequently, both a spontaneous zero field magnetization and magnetostriction are generically expected -- neither are prohibited by Neumann's principle.  The ferromagnetism in Mn$_3$Sn is weak, however, because of the weakness of SIA.  This enables the linear magnetostriction to extend over a very wide field range, simply because of the extended linearity of the magnetization due in turn to the large antiferromagnetic exchange $J$.  Nevertheless, it is to be expected that  residual ferromagnetism accompanies the presence of $\Lambda$. 

\textbf{Piezomagnetism and linear magnetostriction-} Our magnetostriction data should match the piezomagnetic data \cite{Ikhlas2022Piezomagnetic}. Maxwell relations imply a thermodynamic identity between the results of two probes \cite{Lee1955}:
\begin{equation}
(\frac{d\epsilon}{dH})_\sigma=(\frac{dM}{d \sigma})_H
\label{Maxwell}
\end{equation}

The left hand side is the slope of field-linear magnetostriction and the right hand side is the piezomagnetic coefficient. Both represent the components of the same third-rank tensor $\Lambda_{ijk}$ \cite{Borovik-romanov1994}.  Measuring the stress dependence of magnetization, Ikhlas \textit{et al.} \cite{Ikhlas2022Piezomagnetic} reported that  $\frac{dM}{d\sigma}\simeq -0.055~ $Gauss MPa$^{-1}$ for the $x$-axis orientation (and slightly smaller in another sample along the $y$-direction). Now, magnetization in S.I. units is expressed in A/m and  their results corresponds to $\frac{dM}{d\sigma}\simeq -4.4 \times10^{-6}$ T$^{-1}$ in S. I. units. This is significantly smaller that what  we found in four different samples for both longitudinal and transverse magnetostriction. The absolute amplitude of $\frac{\partial \epsilon}{\partial B}$  was always larger  than $ 1 \times10^{-5}$ T$^{-1}$). 

However, as seen in  Fig.\ref{different samples}(d) the discrepancy fades away by considering the fact that the spontaneous magnetization in the sample subject to the piezomagnetic study was smaller than our samples, indicating that its stoichiometry was different from ours.   The calculated piezomagnetic coefficient of Mn$_3$Sn$_{1-\delta}$ is in the same order of magnitude as the experimental results(see supplementary Figure 10 in the supplementary materials \cite{SM}). However, any quantitative account of $\Lambda$ should explain the linear proportionality between piezomagnetic/magnetostrictive response and the spontaneous magnetization remains unexplained.  This requires to consider the role of Sn vacancies.

\textbf{Unoccupied Sn vacancy sites-} Mn$_3$Sn crystal have a deficit of Sn atoms \cite{KREN1975226,Ikhlas2020sampleb}. A number of Sn sites are vacant. As seen in  Fig.\ref{different samples}(b), the amplitude of spontaneous magnetization increases with increasing density of these Sn vacancies. Moreover, the spontaneous magnetization calculated by first-principle theory can be perfectly matched with our experimental data. In contrast, if vacant Sn sites were occupied by Mn atoms (giving rise to a chemical formula of Mn$_{3+x}$Sn$_{1-x}$) the calculated  spontaneous magnetization is two orders of magnitude larger than the experimental data (see supplementary Figure 9 in the supplementary materials \cite{SM}). Note that this calculation was performed for T= 0 K (See More Computational Details in the supplementary materials \cite{SM})).  The agreement between theory and experiment leads to conclude that the chemical formula is Mn$_3$Sn$_{1-\delta}$.

\begin{figure}
\includegraphics[width=5.2cm]{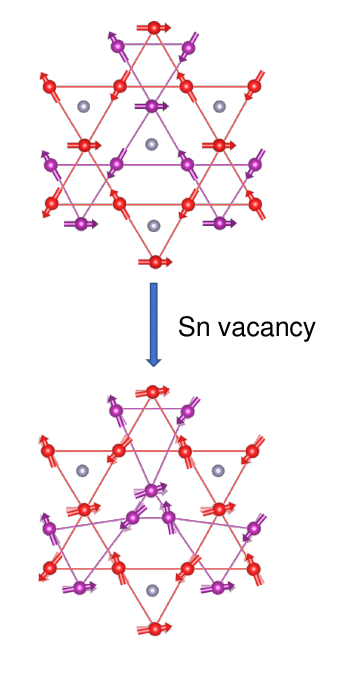}
\caption{\textbf{Possible outcome of a Sn vacancy.}
The vacancy of Sn atoms may generate residual ferromagnetism by twisting the spins of neighboring regular Mn atoms.}
 \label{local spins}
 \end{figure}

One expects two important effects introduced by these Sn vacancy sites. First,  they disrupt the local antiferromagnetic spin cancellation, and induce spin twisting and a local magnetic moment (Fig.\ref{local spins}). Second, they break the high symmetry of the nearby Mn clusters, which may change the orbital character of these moments, enhancing spin-orbit effects and SIA.  These two effect provide a basis for understanding  the observed trends.  Spontaneous magnetization increases with non-stoichiometry because each Sn vacancy brings an uncompensated moment.  The  coupling $\gamma_2$, which was shown in Eq.\ref{eq:45} to drive to $\Lambda \propto M_s$, can only arise from enhanced SIA.  These observations are strongly suggestive. A quantitative account of $\Lambda$ in this picture, requires a Hamiltonian for the non-stoichiometric compound, a substantial theoretical effort, beyond the scope of the present work. 

\textbf{Twisto-magnetic stripes-} In the previous section, we saw that at a small field of $B \simeq 0.02 $ T, longitudinal magnetostriction shows a discontinuous drop but the transverse magnetization shows just a peak. At this field,  the spontaneous magnetization begins to change sign, signaling that the sample is no more single domain \cite{Li2018,Li2019}. The accumulated magnetic energy (i.e. the product of magnetization and magnetic field) at $B_0$ becomes sufficiently large to pay the energy cost of building domain walls. 

As seen in Fig. \ref{lowB}(a) and (b), at $B = B_0$, longitudinal magnetostriction falls to zero. The total  volume of the sample is therefore occupied by magnetic domains whose mutual mechanical deformation along the orientation of the magnetic field cancel each other. Since the total magnetization  is not zero at $B_0$, the multi-domain regime, restricted to a field window between  0.02 T to 0.2 T, there should be a concomitant modulation of lattice parameter and spin orientation (Fig. \ref{lowB}(e)) at Mn sites. When two neighboring atomic triangles have slightly different lattice parameters, atomic-bond angles and spin twisted angles, the force exerted by the strain gradient will be compensated by the magnetic force due to the angle between twisted spins (see Fig.\ref{lowB}(f)). Given that the magnetic energy density (that is the product of magnetization and magnetic field) is only a few pascals \cite{Li2018}), which is orders of magnitude smaller than the elastic energy density, the differential angle of neighboring spins should be orders of magnitude larger than the differential angle of neighboring atomic bonds. A magnetic field of 13T can distort the spin structure by around two degrees\cite{Li2022torque}. 
This intricate mesoscopic structure emerges as a subject of futures investigation by experimental probes in both real space and in momentum space.

\begin{table}[h]
\begin{tabular}{|c|c|c|c|c|}
\cline{1-5}
\multirow{2}{*}{Materials}  &  $\Lambda_{ij}$ & $T$ & \multirow{2}{*}{PM or LM}  & \multirow{2}{*}{Refs}\\
 & ($10^{-6}$ T$^{-1}$) & (K) &   & \\
\cline{1-5}
 
\multirow{3}{*}{CoF$_2$ } & $\Lambda_{14}$ = 21 & 20 & PM & \cite{borovik2013piezomagnetism} \\
 & $\Lambda_{36}$ = 8.2 & 20 & PM & \cite{borovik2013piezomagnetism} \\
 & $\Lambda_{36}$ = 9.8 & 4 & LM & \cite{borovik2013piezomagnetism} \\
\cline{1-5}
 
 \multirow{2}{*}{MnF$_2$ } & $\Lambda_{14}$ = 0.2 & 20 & PM & \cite{borovik2013piezomagnetism} \\
  & $\Lambda_{36}$ = 0.07 & 60 & PM & \cite{baruchel1988piezomagnetism} \\
\cline{1-5}

DyFeO$_3$ & $\Lambda_{36}$ = 6 & 6 & LM & \cite{borovik2013piezomagnetism} \\
\cline{1-5}

YFeO$_3$ & $\Lambda_{15}$ = 1.7 & 6 & LM & \cite{borovik2013piezomagnetism}\\
\cline{1-5}

YCrO$_3$ & $\Lambda_{15}$ = 1 & 6 & LM & \cite{borovik2013piezomagnetism}\\
\cline{1-5}

Terfenol-D & $\Lambda_{11}$ = 4000 & 300 & LM & \cite{Sandlund1994}\\
\cline{1-5}

\multirow{6}{*}{$\alpha$-Fe$_2$O$_3$}  & $\Lambda_{22}$ = 3.2 & 77 & PM &\cite{borovik2013piezomagnetism} \\
& $\Lambda_{22}$ = 1.9 & 78 & LM & \cite{borovik2013piezomagnetism}\\
& $\Lambda_{22}$ = 1.3 & 100 & LM & \cite{borovik2013piezomagnetism}\\
  & $\Lambda_{14}$ = 1.7 & 77 & PM &\cite{borovik2013piezomagnetism} \\
 & $\Lambda_{14}$ = 0.3 & 78 & LM & \cite{borovik2013piezomagnetism}\\
 & $\Lambda_{14}$ = 0.9 & 10 & LM & \cite{borovik2013piezomagnetism}\\
\cline{1-5}

UO$_2$ & $\Lambda_{14}$ = 10.5 & 2.5 & LM &  \cite{Jaime2017} \\
\cline{1-5}

\multirow{2}{*}{
Mn$_{3}$Sn$_{1-\delta}$
} &  $\Lambda_{11}$ = 4.4 & 300 &  PM & \cite{Ikhlas2022Piezomagnetic} \\
&  $\Lambda_{22}$ = 3.8 & 300 &  PM & \cite{Ikhlas2022Piezomagnetic} \\
\cdashline{1-5}
\multirow{3}{*}{
Mn$_{3}$Sn$_{0.891}$
} & $\Lambda_{11}$ = 14.6 & 300 & LM & this work \\
 & $\Lambda_{22}$ = 15.4 & 300 & LM & this work \\
 & $\Lambda_{12}$ = 11.1 & 300 & LM & this work \\
\cline{1-5}

\end{tabular}
\caption{\textbf{$\Lambda_{ij}$ in several magnetically-ordered solids.} Note the limited range of variety in amplitude among solids with different ordering temperatures. Note also that there are cases of discrepancy between measurements of piezomagnetism and magnetostriction.  }
\label{table_comparison_LM-PM}
\end{table}

\textbf{Comparison with other magnets} Table \ref{table_comparison_LM-PM} lists solids in which linear magnetostriction and piezomagnetism have been observed  \cite{Borovik-romanov1994}. We saw that in the case of Mn$_3$Sn, the departure from stoichiometry plays a crucial rule in setting the amplitude of $\Lambda$. Notice the discrepancy in the reported amplitudes of linear magnetostriction and piezomagnetism in other cases, which is yet to be sorted out. Measuring the same sample with both methods could settle the issue.

In summary, we find that  Mn$_3$Sn hosts a large in-plane magnetostriction dominantly linear in magnetic field. We argue that it arises from the field-induced twist of spins and its amplitude can be accounted for given the magnetic and the elastic energy scales. Thermodynamic consistency between the magnetostriction and the piezomagnetic data is achieved only by considering the fact that both, as well as spontaneous magnetization  depend on the concentration of Sn vacancies. The sudden vanishing of longitudinal magnetostriction at the onset of domain nucleation implies the existence of twisto-magnetic stripes with a concomitant modulation of strain and local magnetization in a narrow field window.

%This work was supported by The National Key Research and Development Program of China (Grant No.2022YFA1403500), the National Science Foundation of China (Grant No.12004123, 51861135104 and No.11574097) and  the Fundamental Research Funds for the Central Universities (Grant no. 2019kfyXMBZ071). K. B. was supported by the Agence Nationale de la Recherche (ANR-19-CE30-0014-04). S. J. acknowledges a PhD scholarship by the China Scholarship Council (CSC). X. L. was supported by The National Key Research and Development Program of China (Grant No.2023YFA1609600) and the National Science Foundation of China (Grant No. 12304065).  L.B. was supported by the NSF CMMT program under Grant No. DMR-2116515.

\noindent
* \verb|lixiaokang@hust.edu.cn|\\
* \verb|zengwei.zhu@hust.edu.cn|\\
*\verb|kamran.behnia@espci.fr|\\
%\bibliographystyle{unsrt}
%\bibliography{ref.bib}

\textbf{Methods:}

\textbf{Sample}
%\section*{Methods}

The Mn$_3$Sn single crystals used in this work were grown using the vertical Bridgman method~\cite{Li2018}. The millimeter-size samples were cut from the as-grown samples by a wire saw.

Crystal orientations were determined by a single crystal XRD diffractometer (XtaLAB mini II of Rigaku). The variation of temperature during magnetostriction measurements remained within 0.05\%. 

\textbf{Measurement}

A capacitive dilatometer with a resolution of $\Delta L=0.02$ {\AA} made by Innovative Measurement Technology was used to measure magnetostriction in an Oxford Instruments Teslatron PT. We used an Andeen-Hagerling 2550A capacitance bridge for our measurements. By installing the dilatometer perpendicular to the magnetic field, we also measured magnetostriction in the transverse configuration.  The reliability of our set-up was checked by measuring magnetostriction in a cobalt single crystal (see the supplementary materials \cite{SM}).

\textbf{Calculation}

The theoretical computation was realized by first principles calculations using Vienna ab initio simulation package (VASP)\cite{Kresse1993} with the projector augmented wave (PAW) pseudopotential\cite{Bl1994} and the Perdew-Burke-Ernzerhof (PBE) type of the generalized gradient approximation (GGA) of the exchange correlation potential\cite{perdew1996}. See More Computational Details in the supplementary materials\cite{SM}.

\textbf{Data availability:}
The data that support the findings of this study are available from the corresponding author upon reasonable request. Source data are provided with this paper.

\textbf{Acknowledgements:}
This work was supported by The National Key Research and Development Program of China (Grant No. 2023YFA1609600, 2022YFA1403500), the National Science Foundation of China (Grant No. 12304065, 12004123, 51861135104 and 11574097) and the Fundamental Research Funds for the Central Universities (Grant No. 2019kfyXMBZ071). K. B. was supported by the Agence Nationale de la Recherche (ANR-19-CE30-0014-04). S. J. acknowledges a PhD scholarship by the China Scholarship Council (CSC). L.B. was supported by the NSF CMMT program under Grant No. DMR-2116515.

\textbf{Author contributions statement:}
X.L., Z.Z., and K.B. conceived and designed the study. Q.M. helped by P.N, L.X., J. W., S.J., and H.Z. performed the measurements. J.D. and J.Z. performed the $ab initio$ calculations. L.B. carried out a theoretical analysis of the linear magnetostriction. Q.M., X.L., Z.Z., L.B., and K.B. analyzed the data. Q.M., X.L., Z.Z., L.B., and K.B. wrote the manuscript with assistance from all the authors.

\textbf{Competing interests:}
The authors declare no competing interests.

\textbf{Additional information:}
Correspondence and requests for materials should be addressed to Xiaokang Li, Zengwei Zhu and Kamran Behnia.

\normalem

%\appendix
%
%\begin{center}{\large\bf appendixal Material for ``Magnetostriction, piezomagnetism and domain nucleation in Mn$_3$Sn"}\\
%\end{center}

% ****** Start of file apssamp.tex ******
%
%   This file is part of the APS files in the REVTeX 4.2 distribution.
%   Version 4.2a of REVTeX, December 2014
%
%   Copyright (c) 2014 The American Physical Society.
%
%   See the REVTeX 4 README file for restrictions and more information.
%
% TeX'ing this file requires that you have AMS-LaTeX 2.0 installed
% as well as the rest of the prerequisites for REVTeX 4.2
%
% See the REVTeX 4 README file
% It also requires running BibTeX. The commands are as follows:
%
%  1)  latex apssamp.tex
%  2)  bibtex apssamp
%  3)  latex apssamp.tex
%  4)  latex apssamp.tex
%

%\usepackage[showframe,%Uncomment any one of the following lines to test 
%%scale=0.7, marginratio={1:1, 2:3}, ignoreall,% default settings
%%text={7in,10in},centering,
%%margin=1.5in,
%%total={6.5in,8.75in}, top=1.2in, left=0.9in, includefoot,
%%height=10in,a5paper,hmargin={3cm,0.8in},
%]{geometry}
%\begin{document}

\clearpage

\textbf{Supplementary Materials for ``Magnetostriction, piezomagnetism and domain nucleation in a kagome antiferromagnet"}
\maketitle
%\begin{center}{\large\bf Supplementary Materials for ``Magnetostriction, piezomagnetism and domain nucleation in a kagome antiferromagnet"}\\
%\end{center}

% Add 'S' to the numbering inside the appendix
\renewcommand{\thesection}{Supplementary Note \arabic{section}}
\renewcommand{\tablename}{Supplementary Table}

\renewcommand{\theequation}{S\arabic{equation}}
\setcounter{section}{0}
\setcounter{figure}{0}
\setcounter{table}{0}
\setcounter{equation}{0}

\section{The magnetic crystal class}
From the symmetric point of view,  linear magnetostriction in antiferromagnets is expected in 35 magnetic crystal classes \cite{Birss1963sm}. As is known, the magnetic effects should be considered under the magnetic space (or Shubnikov) groups, which include the space symmetries as well as a two-valued property of the electron spin. As a result, there are 122 three-dimensional magnetic point groups \cite{Birss1963sm}.  There are 31 magnetic point groups of ferromagnetics which can display a much smaller linear magnetostriction than  even-term magnetostriction.  66 magnetic classes including the 31 classes of ferromagnetics can host piezomagnetism \cite{Birss1963sm}. The rest 66 - 31 = 35 of the 59 antiferromagnetic classes can exhibit linear magnetostriction/piezomagnetism. The total 35 antiferromagntic crystal classes are listed in the ref.\cite{Birss1963sm}. 

\begin{figure}
	\includegraphics[width=8.5cm]{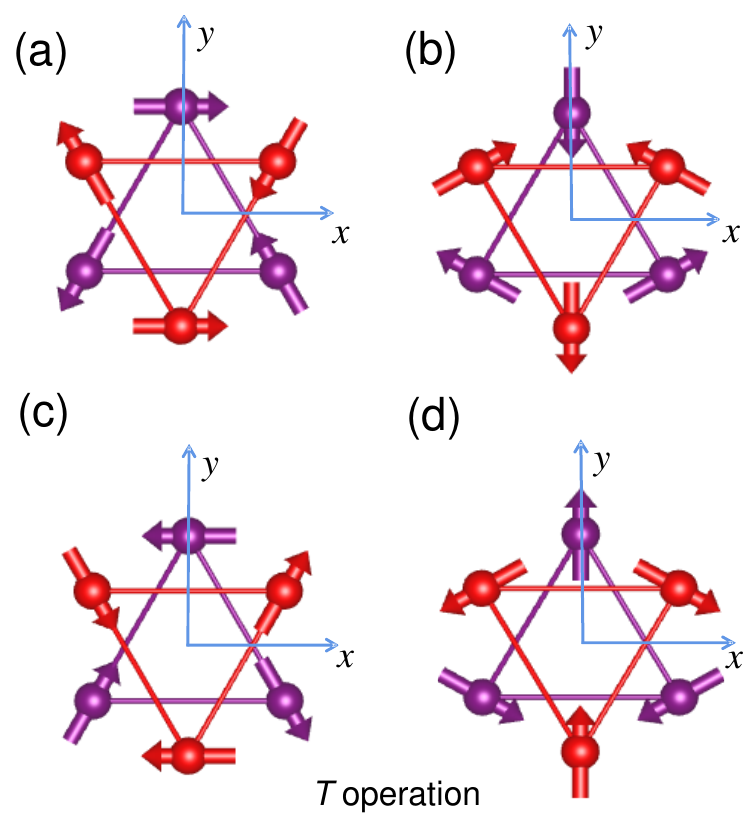}
	\caption{\textbf{Mn spin orientations  after the time-reversal operation} The two configurations (a) and (b) were obtained by experiments as the field is along $x-$ and $y-$ axes \cite{Tomiyoshi1982sm}. The two space groups $Cmc'm'$(a) and $Cm'cm'$(b) belong to the magnetic point group $mm'm'$\cite{Birss1963sm} (see the text for more details).  (c) and (d) show the outcome of $T$ operation on (a) and (b). All spins rotate by 180 degrees.}
	\label{S.Fig:spin_configurations}
\end{figure}

In the Mn$_3$Sn,  Fig. \ref{S.Fig:spin_configurations}(a) shows the spin configuration when the field is along $x$. The symmetry operations are $\{E|0\}$, $\{C_{2x}|0\}$, $T\{C_{2z}|\tau\}$, $T\{C_{2y}|\tau\}$, $\{P|0\}$, $\{PC_{2x}|0\}$, $T\{PC_{2z}|\tau\}$, $T\{PC_{2y}|\tau\}$. $E$, $C$, $T$, $\tau$, and $P$ represent identity, rotation, time-reversal, translation $(0,0, c/2)$ and spatial-inversion operators.  Fig. \ref{S.Fig:spin_configurations} (b) shows the spin configuration when the field is along $y$. The symmetry operations are $\{E|0\}$, $\{C_{2y}|\tau\}$, $T\{C_{2z}|\tau\}$, $T\{C_{2x}|0\}$, $\{P|0\}$, $\{PC_{2y}|\tau\}$, $T\{PC_{2z}|\tau\}$, $T\{PC_{2x}|0\}$. The magnetic space groups of the configurations (a) and (b)  belong to $Cmc'm'$ and $Cm'cm'$, respectively \cite{Suzuki2017sm}. The first letter ``$C$''  represents the type of the Bravais lattice which is base-centered. The next three letters point to the  operations corresponding to the 1st, 2nd and 3rd axis. The $m$, $m'$ and $c'$ represent the symmetric operations: mirror, mirror+time-reversal, mirror+time-reversal+translation along $c$ axis, respectively. These two magnetic space groups are related to magnetic point group $mm'm'$. The list of 35 magnetic classes in  ref.\cite{Birss1963sm}, does not include the magnetic point group $mm'm'$ to which Mn$_3$Sn in its inverse triangular spin state belongs.  

%According to the symmetry group theory, a linear magnetostriction thus should be excluded in Mn$_3$Sn\cite{Birss1963}. The $m,m'$ symmetry does not allow a linear magnetostriction in the current case. 
However, as shown in the Fig. \ref{S.Fig:spin_configurations}(c) and (d) which is the outcome of the $T$ operation of the Fig. \ref{S.Fig:spin_configurations}(a) and (b). As all spins are rotated by 180 degrees, the structure is not same as the started spin texture. Thus, the spin texture has the potential to have linear magnetostriction as a FM state with magnetic point group $mm'm'$. 

\section{Magnetostriction in cobalt}

\begin{figure}
	\includegraphics[width=8.5cm]{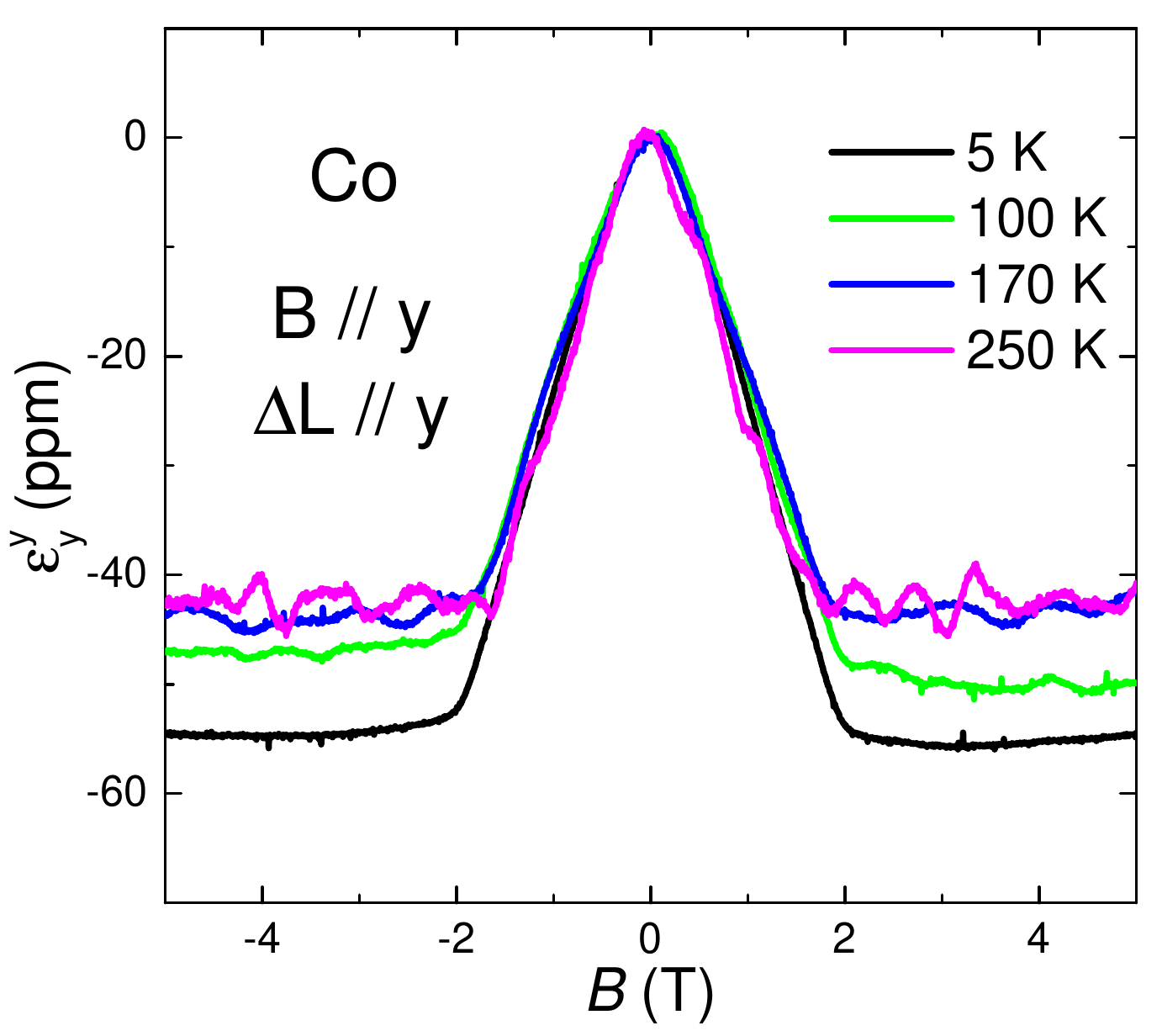}
	\caption{\textbf{Magnetostriction in Cobalt measured by the same setup at different temperatures} For \textit{B}//$\Delta$L//y, The magnetostriction is contractile and tends to saturate above 2 T. }
	\label{S.Co}
\end{figure}

\begin{figure}[htb]
	\includegraphics[width=9cm]{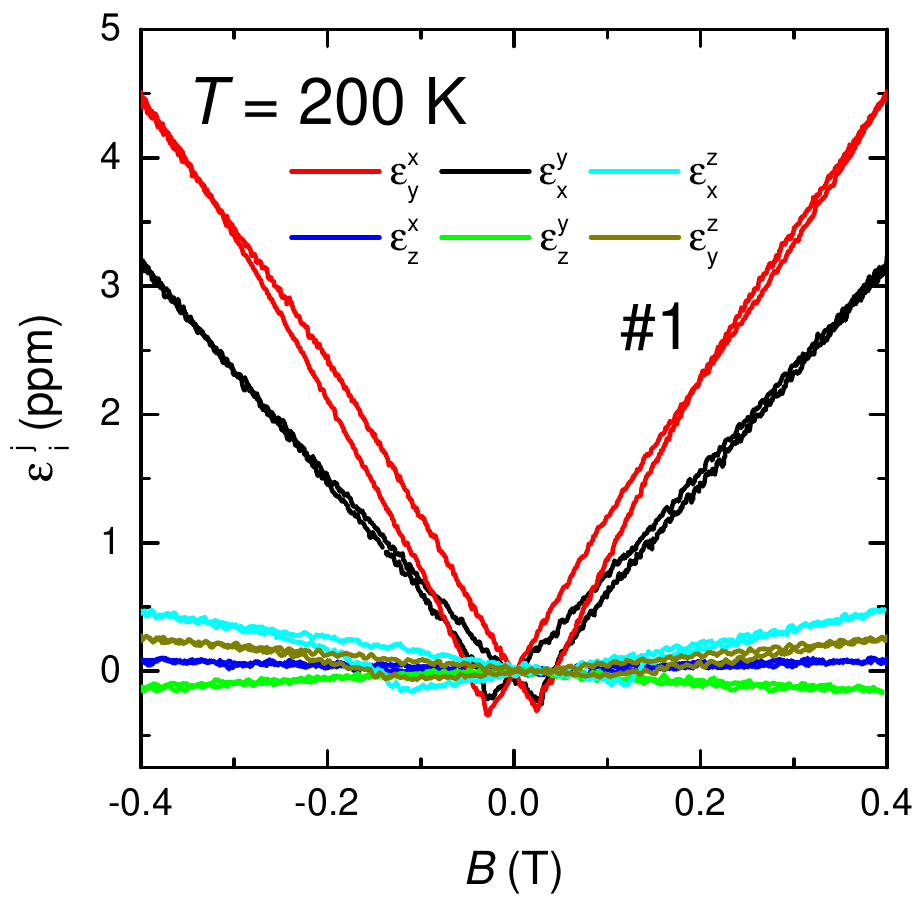}
	\caption{\textbf{Transverse magnetostriction at 200 K in six configurations} The transverse magnetostriction $\epsilon^{z}_x$, $\epsilon^{z}_y$, $\epsilon^{x}_y$,$\epsilon^{x}_z$, $\epsilon^{y}_x$ and $\epsilon^{y}_z$. The results indicate the magnetostsriction occurs solely in-plane, pointing to the rotation of the spins is the source for the magnetostriction. slight residual magnetostrictions in $\epsilon^{x}_z$, $\epsilon^{z}_y$, $\epsilon^{z}_x$ and $\epsilon^{z}_y$, are observed, may due to misalignment of the clamped sample or magnetic field. }
	\label{s.Fig.transverse}
\end{figure}

\begin{figure}
	\includegraphics[width=9cm]{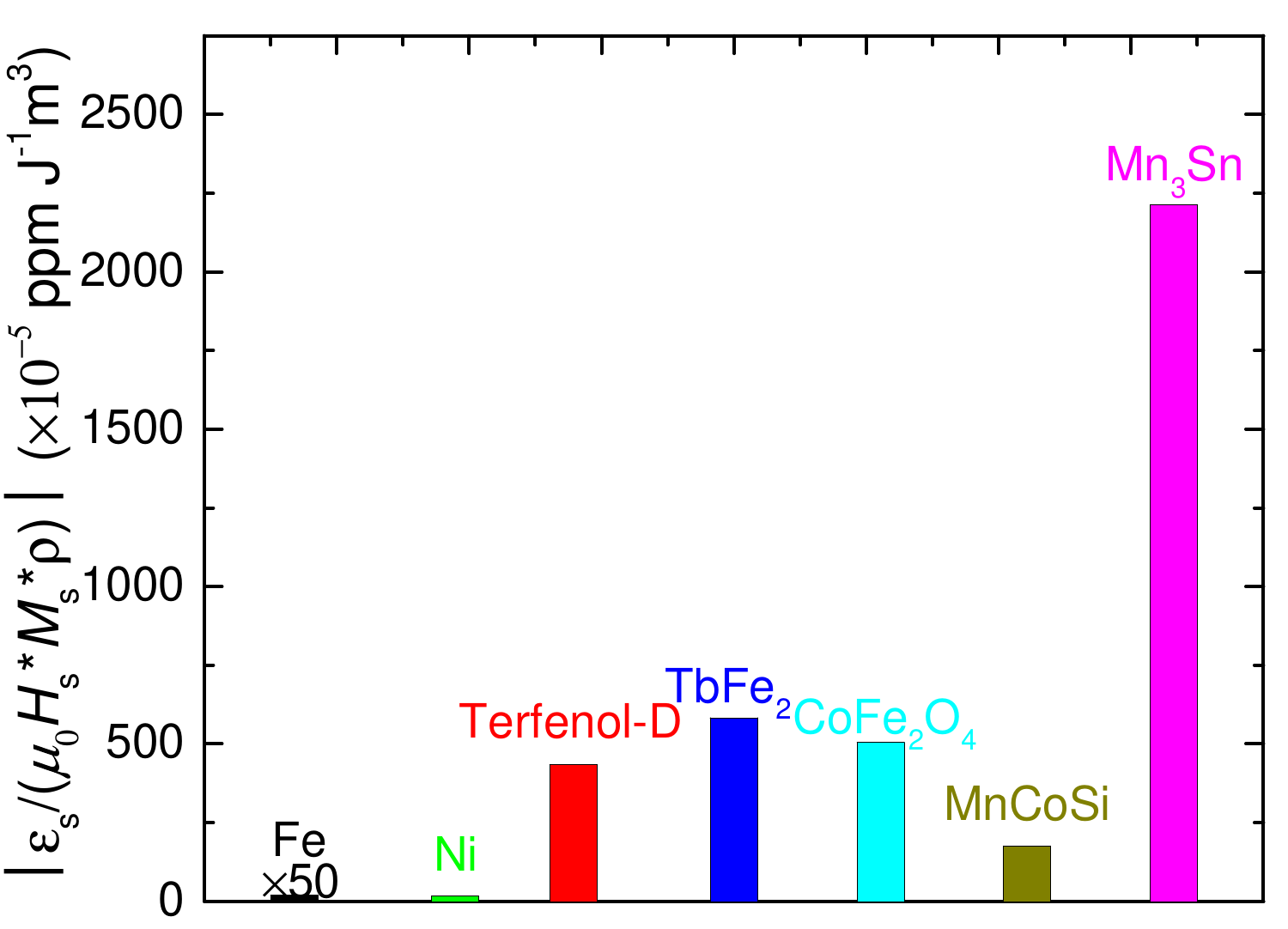}
	\caption{\textbf{Magnetostriction per unit magnetic energy (absolute value) in the several AFM/FM materials.} $H_{s}$ is the saturation field. $M_{s}$ is the magnetization at the saturation field. $\rho$ is the mass density.  $\epsilon_s$ is the magnetostriction coefficient and $\epsilon_s$/($\mu_{0}H_{s}$*$M_{s}$*$\rho$) is the magnetostriction per unit magnetic energy. Mn$_3$Sn stands out by its ratio of magnetostriction to the applied magnetic energy.}
	\label{S_summary}
\end{figure}

\begin{figure}[htbp]
	\includegraphics[width=8.5cm]{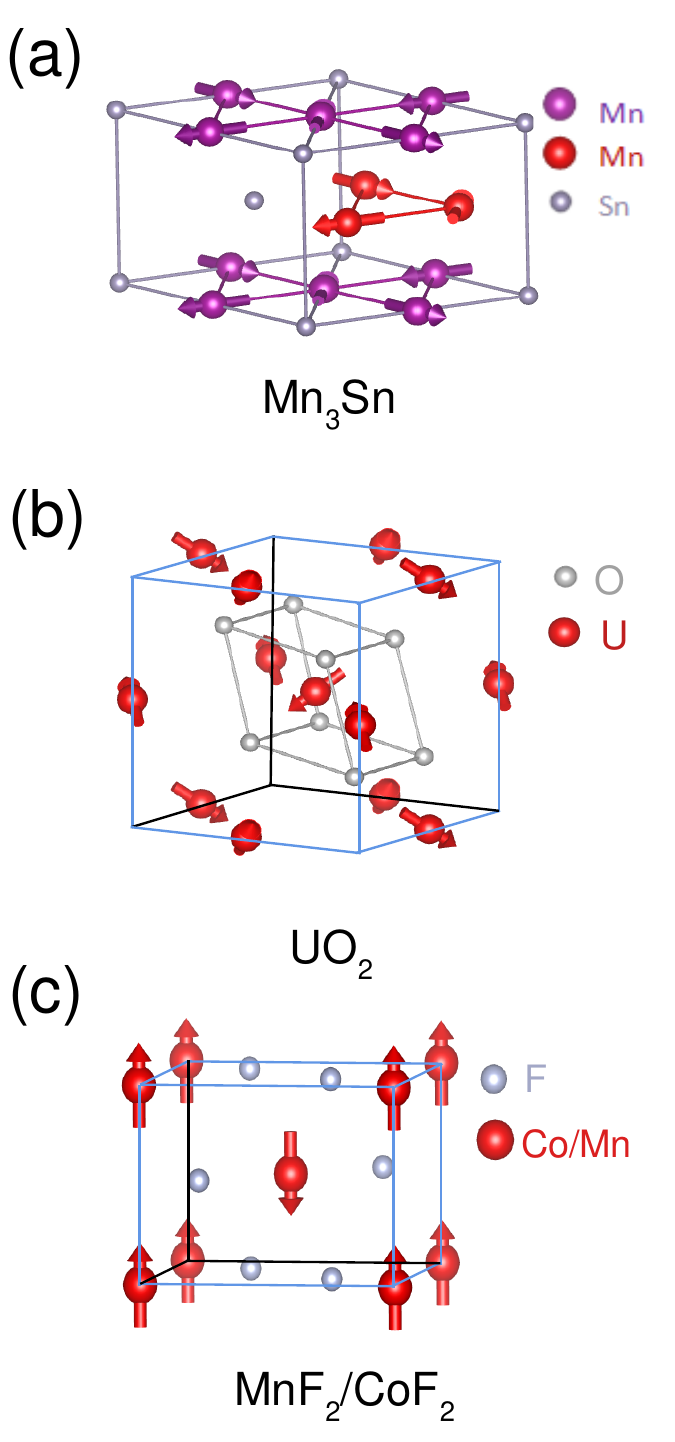}
	\caption{\textbf{Spin textures} Spin configuration in selected antiferromagnets displaying linear magnetostriction: Mn$_3$Sn, CoF$_2$/MnF$_2$ and UO$_2$.} 
	\label{Spin texture}
\end{figure}

%\begin{figure*}
%\includegraphics[width=17cm]{spin texture.jpg}
%\caption{\textbf{Spin textures} Spin configuration in selected antiferromagnets displaying linear magnetostriction: Mn$_3$Sn, CoF$_2$/MnF$_2$ and UO$_2$.} 
%\label{Spin texture}
%\end{figure*}

\begin{figure}[htbp]
	\includegraphics[width=8.5cm]{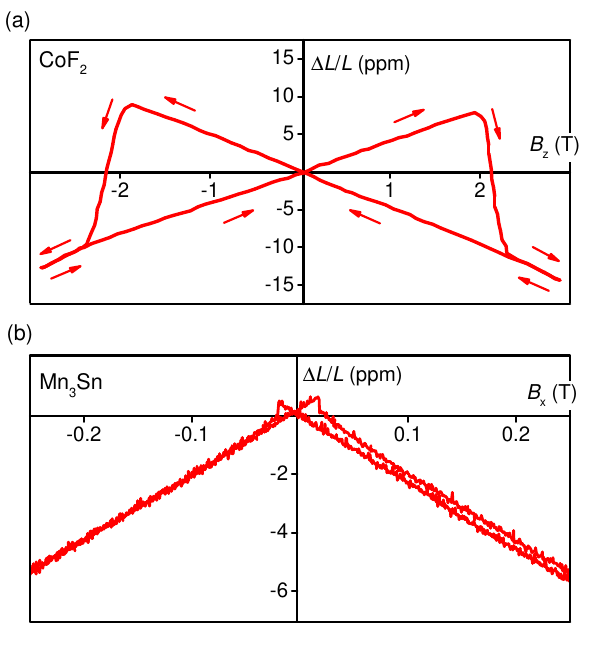}
	\caption{\textbf{Comparison of linear magnetostriction in CoF$_2$ and Mn$_3$Sn} a) The linear magnetostriction of CoF$_2$ with field along $z$ axis\cite{borovik2013piezomagnetismsm}. b) The linear magnetostriction of Mn$_3$Sn with field along $x$ axis. The shape of curves is obvious different.}
	\label{CoF2-Mn3Sn}
\end{figure}

To verify the correctness of the measurement and compare the results in Mn$_3$Sn with a common ferromagnet, we measured the magnetostriction of a commercial cobalt single crystal with dimensions of 2.46 mm$\times$0.568 mm(in-plane)$\times$0.65 mm$(c)$. The results are shown in the Fig. \ref{S.Co}. The measured magnetostriction of cobalt is close to what was previously reported \cite{Bozorth1954sm}.

The field dependence of magnetostriction of a ferromagnetic solid is often dominated by the effect of magnetic field on the orientation of magnetic domains. The sample can expand due to the displacement of  domain boundaries, or contract when these boundaries rotate. In our data is dominated by the latter effect, which leads to a negative magnetostrictive response. The saturation above 2 T is where the magnetization saturates too. 

%can be understood by considering three distinct regimes similar to a magnetization curve. In region I, domain boundaries shift, leading to the expansion of the sample. This field range of the region I is narrow. We have not observed in the cobalt. Since this process is in very small field range. More careful and higher precise measurement is required. The region II follows as field is increased, signaling the end of the boundary displacements and the start of the domain rotational processes. As rotations are under the wary, a gradual contraction will occur. The field range of the region II ends about 2 T in cobalt. The magnetostrictio reaches its saturation as the magnetization is saturated. Beyond the saturation, the forced magnetostriction change a little as the field is further increased .

Thus, magnetostriction in this ferromagnet is totally dominated by the field-induced re-organization of magnetic domains, in contrast with  what is observed in antiferromagnets \cite{Lines1979sm}.

\section{Transverse magnetostriction in all configurations}
Fig.\ref{s.Fig.transverse} shows the transverse magnetostriction for six different configurations: $\epsilon^{z}_x$, $\epsilon^{z}_y$, $\epsilon^{x}_y$, $\epsilon^{x}_z$, $\epsilon^{y}_x$ and $\epsilon^{y}_z$. The transverse magnetostriction is large for both in-plane configurations ($\epsilon^{x}_y$ and $\epsilon^{y}_x$). In contrast, it is   tiny for all four other configurations ($\epsilon^{z}_x$, $\epsilon^{z}_y$, $\epsilon^{x}_z$,  and $\epsilon^{y}_z$). 

The tiny signals in the four out-of-plane  configurations  may be contaminated by misalignment between the clamped sample and magnetic field. Assuming that the field is tilted from the $z$-axis towards the $x$-axis by 5 degrees ( or the sample itself is mistakenly tilted 5 when cut) would generate an erroneous signal of $\epsilon^{x}_x \times$ cos 85$^\circ$ = 8 ppm cos 85$^\circ$ = 0.7 ppm. This is close to the measured value in the $\epsilon^{z}_x$ measurement.  

Thus, the amplitude of out-of-plane transverse magnetostriction remains small and within our experimental margin of error. 

\begin{table*}
	\begin{tabular}{|c|c|c|c|c|c|c|c|}
		\cline{1-8}
		\multirow{2}{*}{Materials} &\multirow{2}{*}{Magnetism}  & $\mu_{0}H_{s}$ & $M_{s}$ & $\rho$ & $\epsilon_s$ &  $\epsilon_s$/($\mu_{0}H_{s}$*$M_{s}$*$\rho$) &\multirow{2}{*}{Refs} \\
		&  & (T) & (emu/g)& (g/cm$^3$) &(ppm)  & ($\times$10$^{-5}$ppmJ$^{-1}$m$^3$)  & \\
		\cline{1-8}
		Fe & FM & 2.15& 217.6 & 7.88 & -14&-0.38&\cite{Crangle1971sm},\cite{Dapino2002sm} \\
		\cline{1-8}
		Ni& FM & 0.61 & 54.68 & 8.9& -50 &-16.8& \cite{Crangle1971sm},\cite{Dapino2002sm}\\
		\cline{1-8}
		%Co\cite{Hall1959} & 1.79& - & - & -93&-&\\
		
		%\hline
		CoFe & FM & 2.45& - & 8.25 &87&-&\cite{Dapino2002sm}\\
		\cline{1-8}
		NiFe & FM & 1.6& - & - & 19 & - &\cite{Dapino2002sm} \\
		\cline{1-8}
		TbFe$_2$ & FM & 1.1& 45 & 9.1 & 2630 & 584 & \cite{Rhyne1974sm},\cite{Dapino2002sm}\\
		\cline{1-8}
		CoFe$_2$O$_4$ & FM & 0.3& 78 & 5.28 & -624 & -505 &\cite{Sukhorukov2017sm} \\
		\cline{1-8}
		Terfenol-D & FM & 0.5& 80 & 9.2 & 1600 & 434.8 &\cite{Clark1978sm},\cite{Abbundi1977sm} \\
		\cline{1-8}
		%YFeO$_3$ & FM & 8& 2.2 & - & -12 & - & \cite{kadomtseva1981direct},\cite{yuan2012effect}\\
		%\hline
		NdCo(30K) & FM & 1& 55 & - & 100 & - &\cite{wang2021magnetocaloricsm}\\
		\cline{1-8}
		YFeO$_3$ & AFM & 8& 2.2 & - & -12 & - & \cite{kadomtseva1981directsm},\cite{yuan2012effectsm}\\
		\cline{1-8}
		Fe$_2$O$_3$ & AFM & 0.04&- & - &64 &-&\cite{urquhart1956magnetostrictivesm},\cite{goyal2012structuralsm}\\
		\cline{1-8}
		ZnCr$_2$Se$_4$(4.3K) & AFM & 6 & - & - & 650 & - & \cite{hemberger2007largesm}\\
		\cline{1-8}
		MnCoSi & AFM & 2.6& 110 & 6.37 & -3200 & -176 &\cite{Gong2015sm} \\
		\cline{1-8}
		Mn$_3$Sn & AFM & 0.02 & 0.13 & 7.47 & 0.43 & 2214 & this work\\
		\cline{1-8}
	\end{tabular}
	\caption{\textbf{Amplitude of longitudinal magnetostriction  and magnetization in different materials}. Unless otherwise specified, the data corresponds to room temperature.  $H_s$ is the threshold magnetic field of domain nucleation, $M_s$ and $\epsilon_s$ are the corresponding magnetization and magnetostriction under this magnetic field respectively. $\rho$ is the mass density.  $\epsilon_s$/($\mu_{0}H_{s}$*$M_{s}$*$\rho$) represents the magnetostriction normalized by the magnetic energy density. The mass density $\rho$ is used for converting magnetization units from emu/g to emu/cm$^3$.
		%The Fe$_2$O$_3$, ZnCr$_2$Se$_4$, MnCoSi and Mn$_3$Sn are AFM, while others are FM materials.
	}
	\label{S_table_comparison}
\end{table*}

\section{Comparison with other  magnetic solids}
The amplitude of magnetostriction in Mn$_3$Sn is not exceptionally large compared to other magnetic materials. However, considering $\epsilon_s$/($\mu_{0}H_{s}$*$M_{s}$*$\rho$), which represents the amplitude of magnetostriction normalized by the magnetic energy density, we can see that this ratio is large. Here,  $H_s$ is the threshold magnetic field of domain nucleation, $M_s$ and $\epsilon_s$ are the corresponding magnetization and magnetostriction under this magnetic field respectively. Table \ref{S_table_comparison} compares longitudinal magnetostriction and magnetization in several magnetic  materials at room temperature (unless otherwise specified).  As seen in Fig.\ref{S_summary},  $\epsilon_s$/($\mu_{0}H_{s}$*$M_{s}$*$\rho$) in Mn$_3$Sn is much larger than other solids. 

Table \uppercase\expandafter {\romannumeral2}  shows piezomagnetic(PM) coefficients $\Lambda$ of selected bulk materials. The $\Lambda_{ijk}$ was deduced either by piezomagnetism (PM) or linear magnetostriction (LM) at specific temperatures.  As seen from the table, the measured $\Lambda_{ijk}$ is different in Fe$_2$O$_3$ and the $\Lambda_{ijk}$  deduced from PM and LM is  different by 1.68($\Lambda_{22}$) and 5.67($\Lambda_{14}$) times, respectively. Thus, the $\Lambda_{ijk}$ difference of 2.73 times between PM and LM should be reasonable. We also note that the piezomagnetic(PM) coefficients $\Lambda_{11}$ is sample dependent (see discussion below). 

Fig.\ref{Spin texture} compares the spin textures of several antiferromagnets in which linear magnetostrition has been detected. The list includes Mn$_3$Sn, CoF$_2$/MnF$_2$ and UO$_2$. The spins in Mn$_3$Sn and UO$_2$ are non-collinear, while CoF$_2$/MnF$_2$ is collinear. We note that CoF$_2$ and MnF$_2$ have been recently identified as altermagnetic candidates \cite{Smejkal2022sm}. 
\begin{figure*}[h]
	\includegraphics[width=17cm]{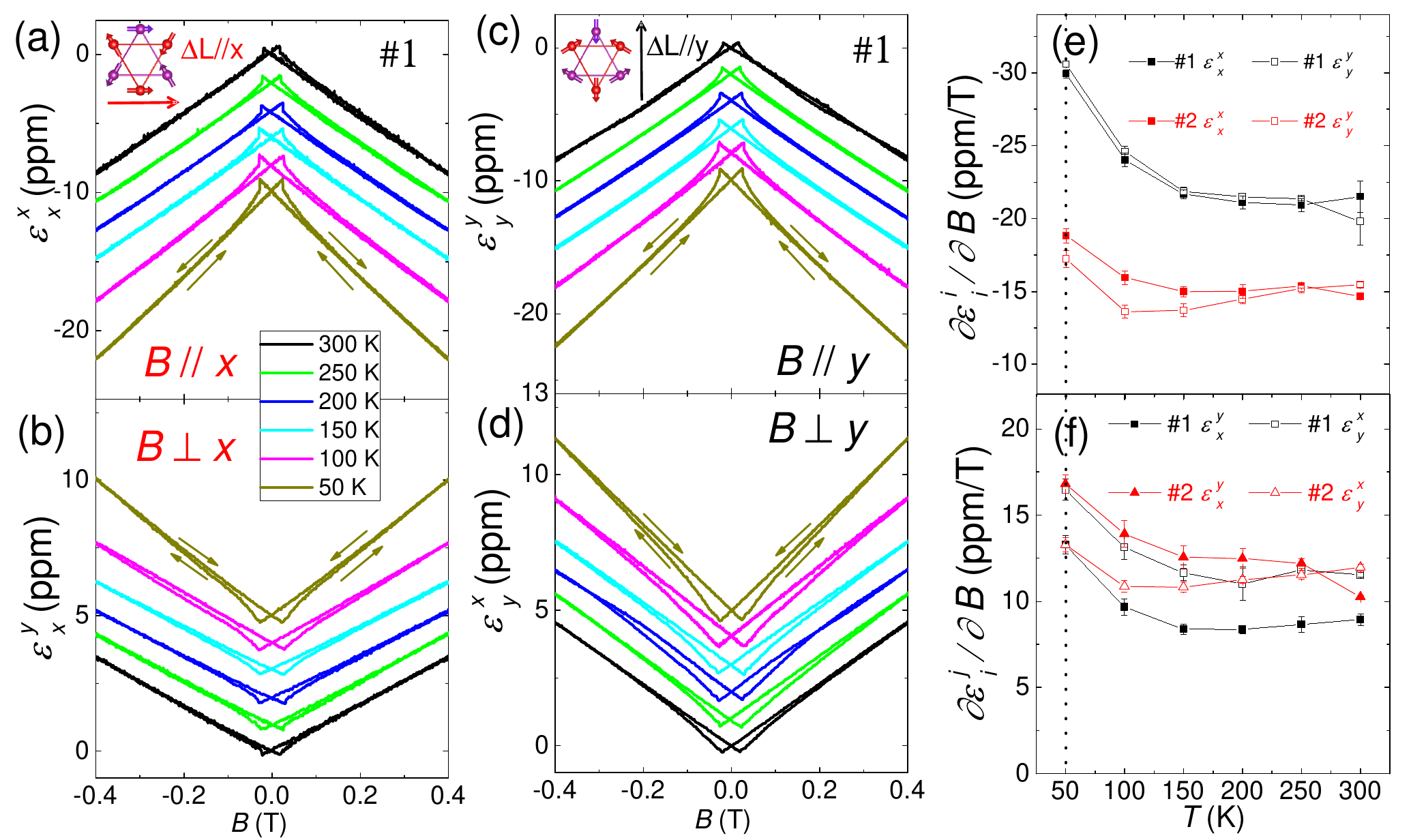}
	\caption{\textbf{Reproduciblity} a-d) Longitudinal and magnetostriction data for sample \#1. To be compared with Fig. 3 in the main text. The hysteresis loop and  linearity of magnetostriction seen in sample \#2 is verified in  sample \#1. (e-f) comparison of the temperature dependence in the two samples.}
	\label{S.sample2}
\end{figure*}

\begin{figure}
	\includegraphics[width=8.5cm]{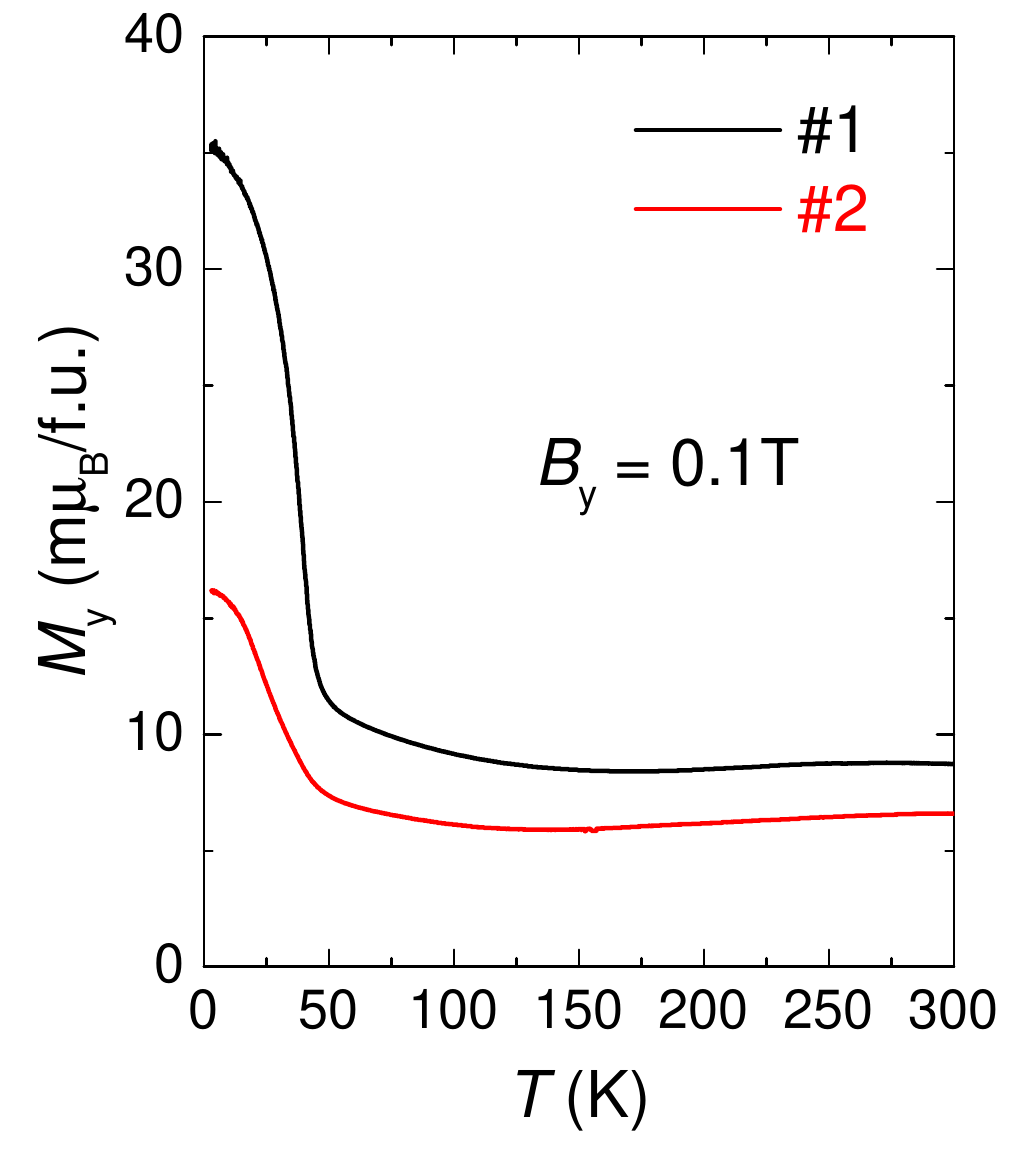}
	\caption{\textbf{Magnetization in samples \#1 and \#2.}  Th temperature dependence of magnetization in two samples showing a difference in amplitude and the temperature below which the triangular order is destroyed.}
	\label{magnetization}
\end{figure}
\begin{figure}
	\includegraphics[width=8.5cm]{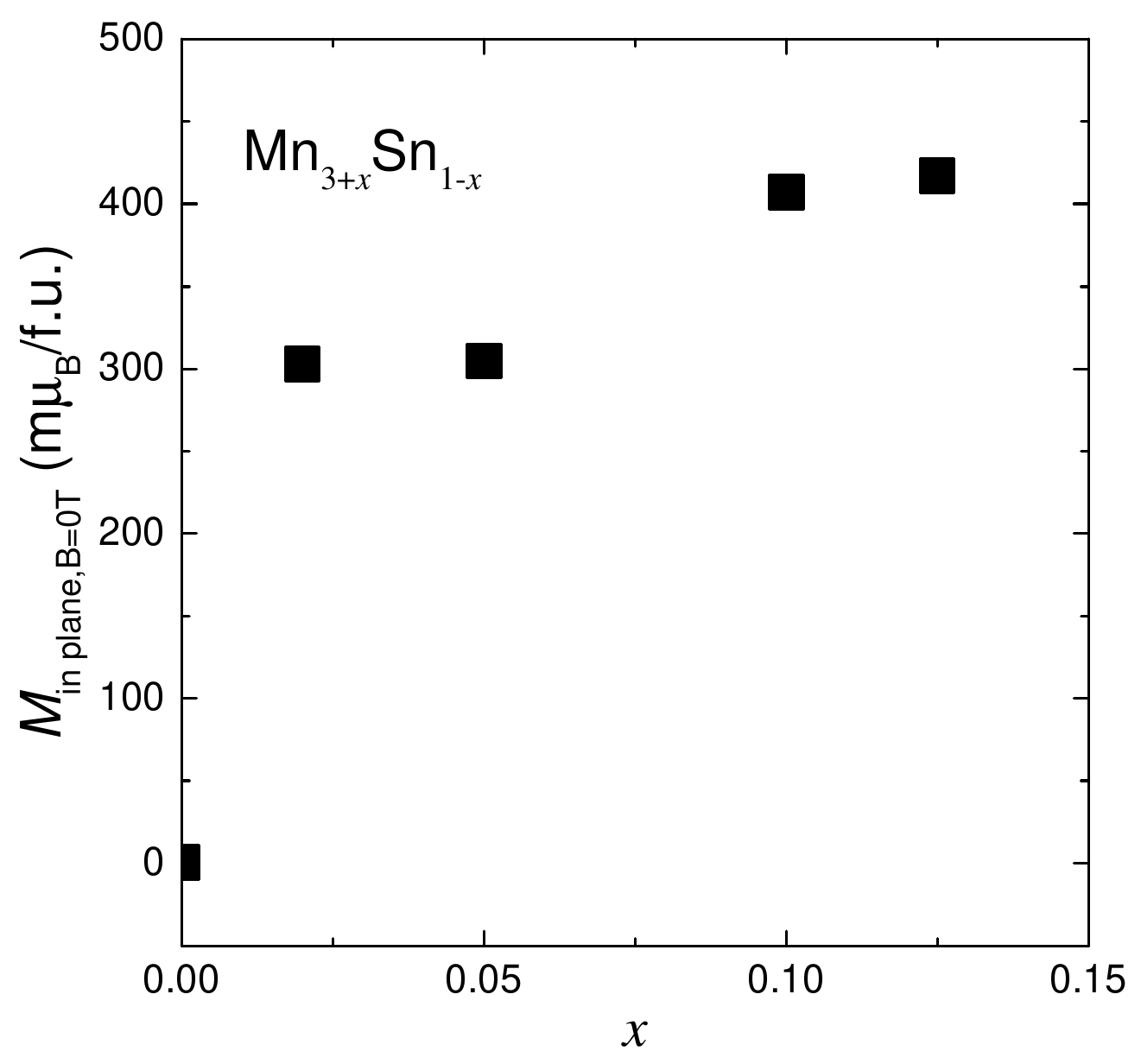}
	\caption{\textbf{The calculated spontaneous magnetization in the case of Mn atoms occupying Sn sites.} It is bar far larger than the experimentally measured magnetization. This made us to conclude that Sn vacancies are not occupied by Mn atoms.}
	\label{Mn-Sn}
\end{figure}
Fig.\ref{CoF2-Mn3Sn} compares the field dependence of magnetostriction between CoF$_2$ \cite{borovik2013piezomagnetismsm} and Mn$_3$Sn. The coercive field of CoF$_2$ with field along $z-$ axis is much larger than that of Mn$_3$Sn. Two features are remarkable. First of all, linear magnetostriction in Mn$_3$Sn (measured at room temperature) is larger than that of CoF$_2$ (measured at cryogenic temperatures). Second, the hysteresis loop in  CoF$_2$ is dominated by multi-domain response and do not display the abrupt jump at domain nucleation observed in Mn$_3$Sn.

\section{Sample dependence}
Fig.\ref{S.sample2} compares longitudinal and transverse magnetostriction in  two samples at different temperatures. Panels (a), (b), (c) and (d) show $\epsilon_x^x$, $\epsilon_x^y$, $\epsilon_y^y$ and $\epsilon_y^x$ for the sample \#1 at various temperature, respectively. 

Panels (e) and (f) show the magnetostriction coefficients for longitudinal and transverse configurations in the two samples. The difference in longitudinal magnetostriction coefficient is about 20\%. The transverse magnetostriction coefficient is comparable with within 10\%. In sample \#1, the amplitude of longitudinal  magnetostriction is twice larger than the amplitude of transverse magnetostriction. In sample \#2, the difference between longitudinal and transverse responses remain within the experimental margin.  Fig.\ref{magnetization} compares the temperature dependence of magnetization of the two samples at 0.1 T. Not only the amplitude of the magnetization is different, but also the destruction of the triangular spin state occurs at different temperatures. Both point to a difference in stoichiometry between the two samples.

\begin{figure*}
	\includegraphics[width=15cm]{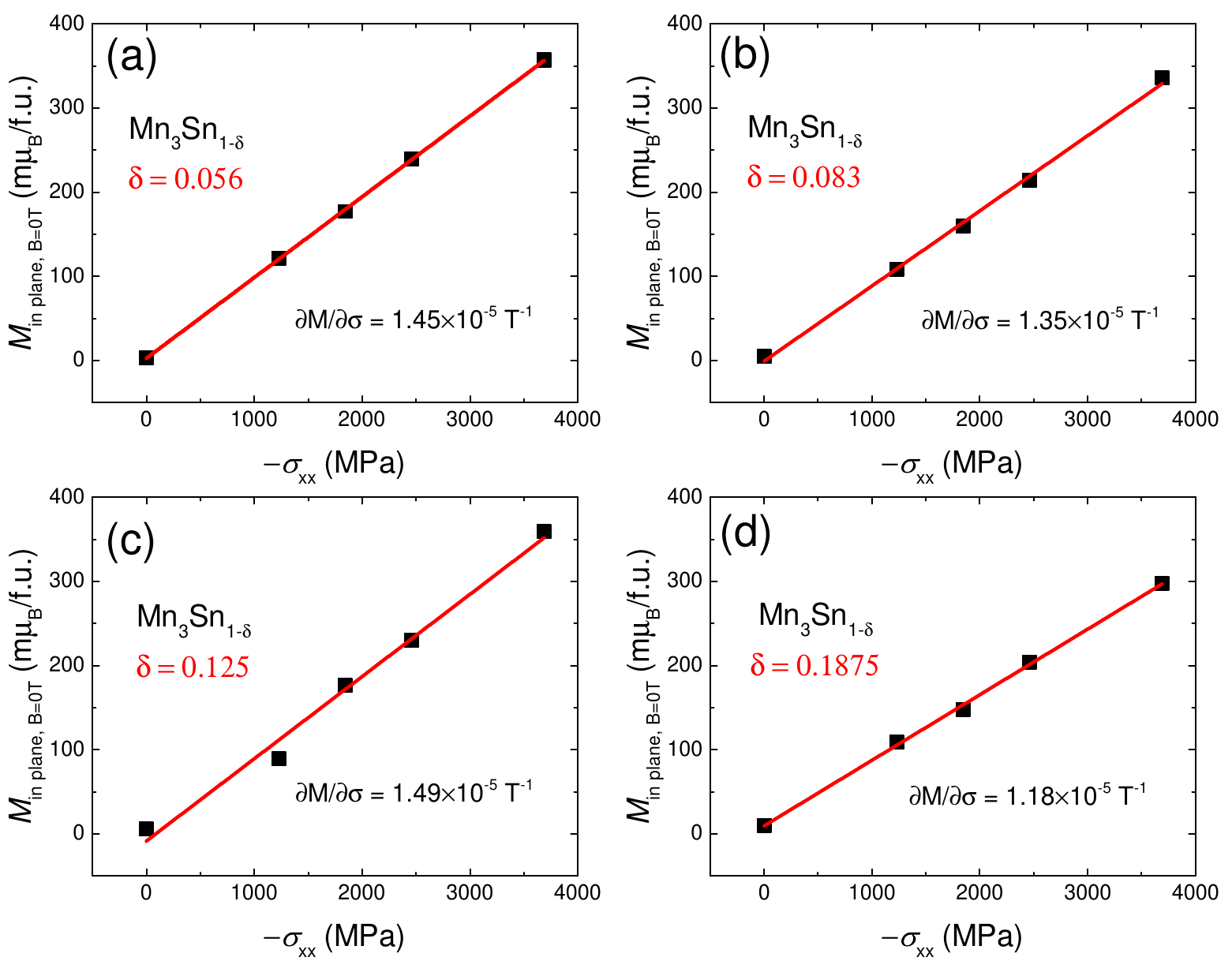}
	\caption{\textbf{The calculated piezomagnetic coefficient in Mn$_3$Sn$_{1-\delta}$.} The order of magneitude agrees with the experimentally measured one. }
	\label{PM}
\end{figure*}
\section{Ab initio calculation}
The theoretical computation was realized by first principles calculations using Vienna ab initio simulation package (VASP)\cite{Kresse1993sm} with the projector augmented wave (PAW) pseudopotential\cite{Bl1994sm} and the Perdew-Burke-Ernzerhof (PBE) type of the generalized gradient approximation (GGA) of the exchange correlation potential\cite{perdew1996sm}. Doping concentration of the bulk Mn$_3$Sn was regulated by the construction of supercell with the vacancy defects at the Sn site. We built supercells containing 4-, 6-, 8- and 9-unit cells for various concentrations of Sn vacancy of 12.5\%, 8.3\%, 5.6\%, and 18.75\%, respectively. Each supercell with one Sn vacancy site except for the concentration of 18.75\%, which has three vacancy sites in the whole supercell. The relative position of the three vacancy sites has been tested to the energy minimal. In terms of seeding the most stable vacancy site, we did not use Monte Carlo method since in the concentration of 12.5\%, 8.3\% and 5.6\%, supercell include only one vacancy site and it is unnecessary to seek between various Sn sites due to the periodic boundary conditions. The lattice constant of Mn$_3$Sn was obtained from experimental data, with $a=b=5.665$ Å and $c=4.531$ Å in the pure configuration, i.e., a state without any vacancy defect. Through first principles calculations, we found that the net magnetization increases with the gradual rise of vacancy concentration (see  Fig.6 b).

On the other hand, in the case of excessive Mn atoms occupying Sn sites, the calculated value of net magnetization is two orders of magnitude larger than what was experimentally measured (see Fig.\ref{Mn-Sn}).

We also calculated the piezomagnetic effect and obtained a linear relationship between the net magnetization and stress. The piezomagnetic coefficient was calculated as follows: $\Lambda=\frac{\partial M}{\partial \sigma}$, where $M$ is the net magnetization, $\sigma$ represents the stress. In addition, to converge the self-consist calculation, an energy cutoff of 500 eV and energy convergence criterion of 10$^{-6}$ eV were adopted with the k-points of 12×12×16 uniformly spread in the Brillouin zone corresponding to the lattice constant we have just mentioned. 

Fig.\ref{PM} shows the results of our calculations of the piezomagnetic effect. The calculated piezomagnetic coefficient is of the same order of magnitude of the experimentally measured one ($\sim$ 10$^{-5}$T$^{-1}$).

\section{Longitudinal magnetostriction at 25K}

\begin{figure*}
	\includegraphics[width=8.5cm]{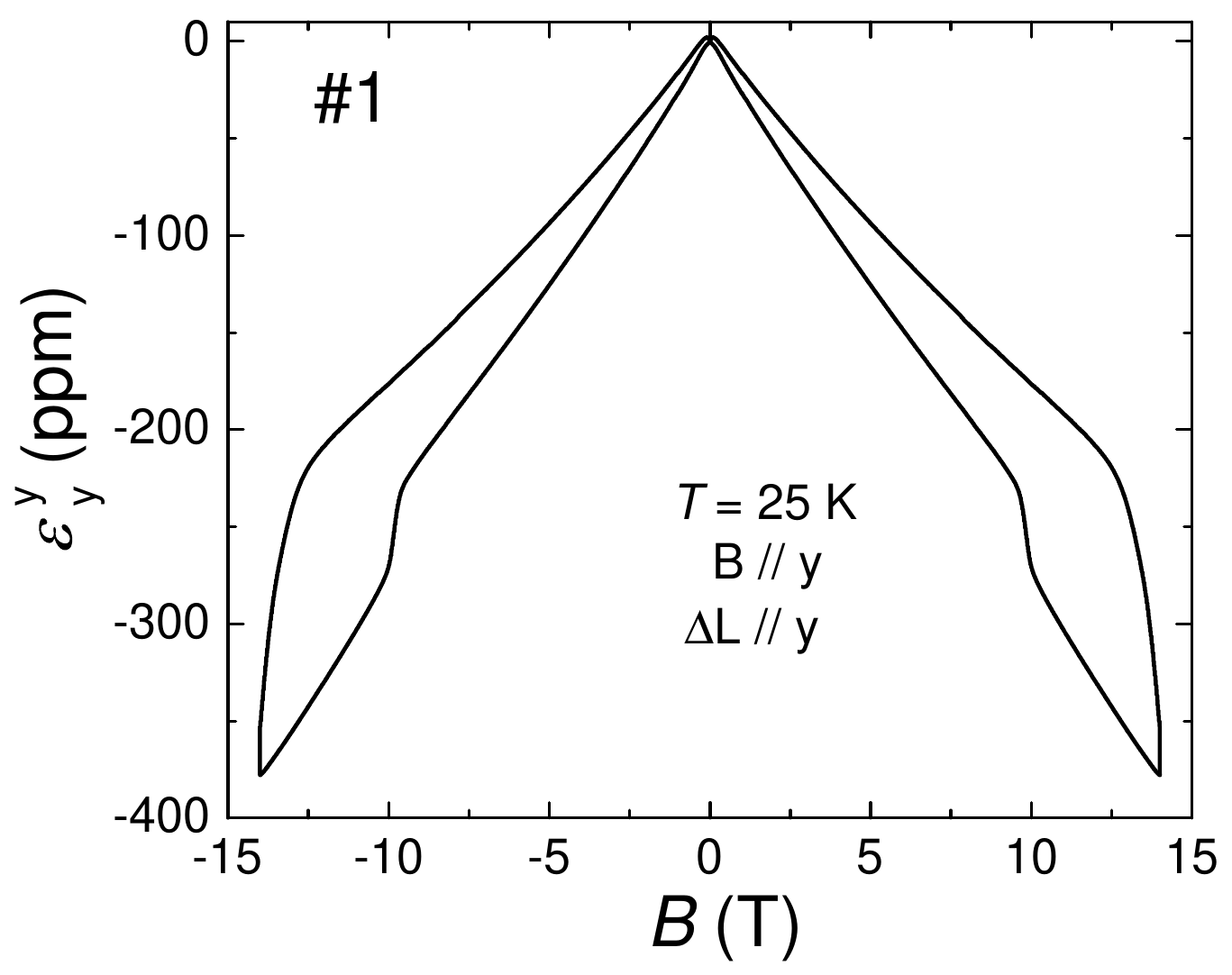}
	\caption{\textbf{Longitudinal magnetostriction $\epsilon_{y}^{y}$ at 25K in sample $\#1$.}  Note the large hysteresis, absent in the high temperature data.}
	\label{25K}
\end{figure*}

Fig.\ref{25K} shows $\epsilon_y^y$ in sample $\#$1 at a temperature of 25K. There is shows a contraction behavior as seen for temperatures exceeding 50 K. However, there is  a very large hysteresis, quite different from the results at high temperatures (50K to 300K), indicating that in the spin glassy phase stabilized below 50 K, rotation of magnetic domains may play a role in the magnetostrictive response.

\section{Correlation between lattice parameters and magnetostriction in different samples}

\begin{figure*}
	\includegraphics[width=17cm]{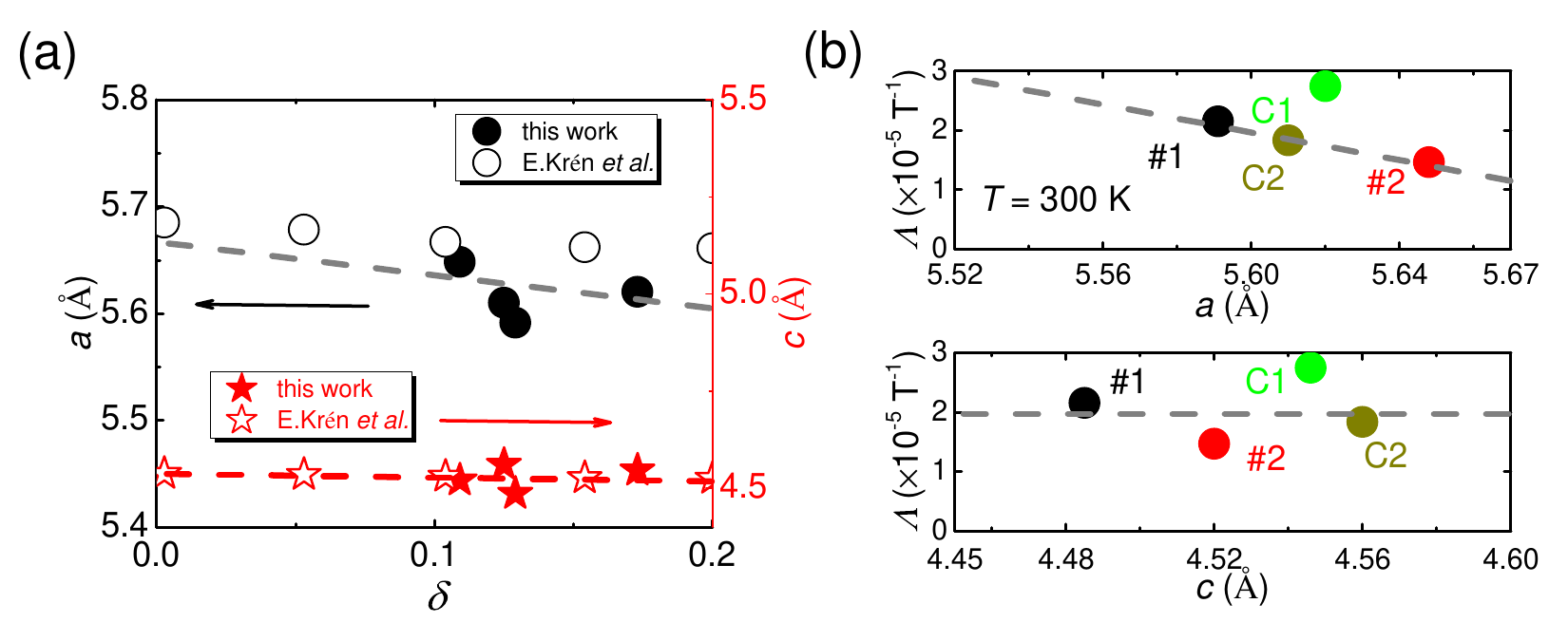}
	\caption{\textbf{Lattice parameters and $\Lambda$ in different samples.} (a) Lattice parameters a and c as a function of the concentration of Sn vacancies. (b) Evolution of $\Lambda$ with lattice parameters a and c, respectively. }
	\label{Lattice parameters}
\end{figure*}

The lattice parameters were measured by a single crystal XRD diffractometer (XtaLAB mini II of Rigaku). Fig.\ref{Lattice parameters}(a) shows the lattice parameters $a$ and $c$ as a function of the Sn vacancy concentration, which is consistent with previous reports\cite{KREN1975226sm}. Fig.\ref{Lattice parameters}(b) shows the variation of piezomagnetic/magnetostriction coefficients with lattice parameters. Note that the difference in lattice parameter among the samples is at least an order of magnitude larger than the change induced by a 10 T magnetic field in each sample.

\clearpage
\normalem

\end{document}